\documentclass[twoside,reqno,12pt]{amsart}
\usepackage{a4wide} 
\usepackage{amsmath}
\usepackage{amsfonts}
\usepackage{amssymb}
\usepackage{listings}
\usepackage{times}
\usepackage{pstricks}

\catcode`\@11\relax
\newif\ify@autoscale \y@autoscaletrue \def\Yautoscale#1{\ifnum #1=0
  \y@autoscalefalse\else\y@autoscaletrue\fi}
\newdimen\y@b@xdim
\newdimen\y@boxdim \y@boxdim=13pt
\def\Yboxdim#1{\y@autoscalefalse\y@boxdim=#1}
\newdimen\y@linethick    \y@linethick=.3pt
\def\Ylinethick#1{\y@linethick=#1}
\newskip\y@interspace \y@interspace=0ex plus 0.3ex
\def\Yinterspace#1{\y@interspace=#1}
\newif\ify@vcenter   \y@vcenterfalse
\def\Yvcentermath#1{\ifnum #1=0 \y@vcenterfalse\else\y@vcentertrue\fi}
\newif\ify@stdtext   \y@stdtextfalse
\def\Ystdtext#1{\ifnum #1=0 \y@stdtextfalse\else\y@stdtexttrue\fi}
\newif\ify@enable@skew   \y@enable@skewfalse
\expandafter\ifx\csname enableskew\endcsname\relax
 \y@enable@skewfalse \else \y@enable@skewtrue\fi
\def\y@vr{\vrule height0.8\y@b@xdim width\y@linethick depth 0.2\y@b@xdim}
\def\y@emptybox{\y@vr\hbox to \y@b@xdim{\hfil}}
\ify@enable@skew
 \def\y@abcbox#1{\if :#1\else
   \y@vr\hbox to \y@b@xdim{\hfil#1\hfil}\fi}
 \def\y@mathabcbox#1{\if :#1\else
   \y@vr\hbox to \y@b@xdim{\hfil$#1$\hfil}\fi}
\else
 \def\y@abcbox#1{\y@vr\hbox to \y@b@xdim{\hfil#1\hfil}}
 \def\y@mathabcbox#1{\y@vr\hbox to \y@b@xdim{\hfil$#1$\hfil}}
\fi
\def\y@setdim{%
  \ify@autoscale%
   \ifvoid1\else\typeout{Package youngtab: box1 not free! Expect an
     error!}\fi%
   \setbox1=\hbox{A}\y@b@xdim=1.6\ht1 \setbox1=\hbox{}\box1%
  \else\y@b@xdim=\y@boxdim \advance\y@b@xdim by -2\y@linethick
  \fi}
\newcount\y@counter
\newif\ify@islastarg
\def\y@lastargtest#1,#2 {\if\space #2 \y@islastargtrue
  \else\y@islastargfalse\fi}
\def\y@emptyboxes#1{\y@counter=#1\loop\ifnum\y@counter>0
  \advance\y@counter by -1 \y@emptybox\repeat}
\def\y@nelineemptyboxes#1{%
  \vbox{%
    \hrule height\y@linethick%
    \hbox{\y@emptyboxes{#1}\y@vr}
    \hrule height\y@linethick}\vskip-\y@linethick}
\def\yng(#1){%
  \y@setdim%
  \hskip\y@interspace%
  \ifmmode\ify@vcenter\vcenter\fi\fi{%
  \y@lastargtest#1,
  \vbox{\offinterlineskip
    \ify@islastarg
     \y@nelineemptyboxes{#1}
    \else
     \y@ungempty(#1)
    \fi}}\hskip\y@interspace}
\def\y@ungempty(#1,#2){%
  \y@nelineemptyboxes{#1}
  \y@lastargtest#2,
  \ify@islastarg
   \y@nelineemptyboxes{#2}
  \else
   \y@ungempty(#2)
  \fi}
\def\y@nelettertest#1#2. {\if\space #2 \y@islastargtrue
  \else\y@islastargfalse\fi}
\def\y@abcboxes#1#2.{%
  \ify@stdtext\y@abcbox#1\else\y@mathabcbox#1\fi%
  \y@nelettertest #2.
  \ify@islastarg\unskip%
   \ify@stdtext\y@abcbox{#2}\else\y@mathabcbox{#2}\fi%
  \else\y@abcboxes#2.\fi}
 \newdimen\y@full@b@xdim
 \newcount\y@m@veright@cnt
\ify@enable@skew
 \def\y@get@m@veright@cnt#1#2.{%
   \if :#1 \advance\y@m@veright@cnt by 1\y@get@m@veright@cnt#2.\fi}
 \let\y@setdim@=\y@setdim
 \def\y@setdim{%
   \y@setdim@ \y@full@b@xdim=\y@b@xdim
   \advance\y@full@b@xdim by 1\y@linethick}
 \def\y@m@veright@ifskew#1{
   \y@m@veright@cnt=0 \y@get@m@veright@cnt#1.
   \moveright \y@m@veright@cnt\y@full@b@xdim}
\else
 \def\y@m@veright@ifskew#1{}
\fi
\def\y@nelineabcboxes#1{%
  \y@nelettertest #1.
  \ify@islastarg
   \y@m@veright@ifskew{#1}
    \vbox{
      \hrule height\y@linethick%
      \hbox{\ify@stdtext\y@abcbox#1\else\y@mathabcbox#1\fi\y@vr}
      \hrule height\y@linethick}\vskip-\y@linethick
  \else
   \y@m@veright@ifskew{#1}
    \vbox{
      \hrule height\y@linethick%
      \hbox{\y@abcboxes #1.\y@vr}%
      \hrule height\y@linethick}\vskip-\y@linethick
  \fi}
\def\young(#1){%
  \y@setdim%
  \hskip\y@interspace%
  \y@lastargtest#1,
  \ifmmode\ify@vcenter\vcenter\fi\fi{%
  \vbox{\offinterlineskip
    \ify@islastarg\y@nelineabcboxes{#1}%
    \else\y@ungabc(#1)%
    \fi}}\hskip\y@interspace}
\def\y@ungabc(#1,#2){%
  \y@nelineabcboxes{#1}%
  \y@lastargtest#2,
  \ify@islastarg\y@nelineabcboxes{#2}%
  \else\y@ungabc(#2)%
  \fi}
\catcode`\@12\relax
 
\allowdisplaybreaks

\newcounter{mycnt}
\def\themycnt{\thesection.\arabic{mycnt}}
\def\mybenv#1{\refstepcounter{mycnt}%
       \vskip 3pt\noindent{\bf #1~~\themycnt}:~}
\def\myeenv{\hfill\rule{1ex}{1ex}\vskip 3pt}
\def\qed{\hfill$\Box$}
\renewcommand{\theequation}{\arabic{section}-\arabic{equation}}
\def\ov{\overline}
\def\1{\ov{1}}
\def\2{\ov{2}}
\def\3{\ov{3}}

\def\nn{\nonumber \\}
\def\Id{\text{I\!d}}

\def\openR{\mathbb{R}}
\def\openC{\mathbb{C}}
\def\openN{\mathbb{N}}

\def\openQ{\mathbb{Q}}
\def\CO{\Delta} 
\def\co{\delta}

\def\antip{{\sf S}}

\def\ip{\star}
\def\outer{m}
\def\inner{\mathsf{m}}
\def\PlethDelta{\nabla}   

\def\la{\langle}
\def\ra{\rangle}

\def\!{\kern -0.15ex}
\def\antip{\textsf{S}}

\def\ip{\star}

\def\cal{\mathcal}
\def\wgt{{\mathrm{wgt}}}

\begin{document}

\title[Plethysms and vertex operators] 
{Plethysms, replicated Schur functions and series, with
applications to vertex operators\footnote{T\lowercase{his research was
supported through the programme} ``R\lowercase{esearch in} P\lowercase{airs}''
\lowercase{by the} M\lowercase{athematisches} F\lowercase{or\-schungsinstitut}
O\lowercase{berwolfach in} 2010.}}
\author{Bertfried Fauser}
\address{%
School Of Computer Science\\
The University of Birmingham\\
Edgbaston, Birmingham, B25 2TT}
\email{b.fauser@cs.bham.ac.uk}
\author{Peter D. Jarvis}
\address{%
University of Tasmania, School of Mathematics and Physics,
GPO Box 252-21, 7001 Hobart, TAS, Australia}
\email{Peter.Jarvis@utas.edu.au}
%
\author{Ronald C. King}
\address{%
School of Mathematics, University of Southampton,
Southampton SO17 1BJ, England}
\email{R.C.King@soton.ac.uk}



\date{June 15, 2010; UTAS-PHYS-2010-09}

\dedicatory{\vskip3ex
\parbox{0.8\textwidth}{
The characters of the orthogonal and symplectic groups have been
found by Schur~\cite{schur:1924a} and Weyl~\cite{weyl:1930a} respectively. 
The method used is transcendental, and depends on integration over the
group manifold. These characters, however, may be obtained by purely
algebraic methods, \ldots. This algebraic method would seem to offer
a better prospect of successful application to other restricted groups
than the method of group integration.\hfill
\vskip2ex
\hfill D.E. Littlewood, Phil. Trans. Roy. Soc. London, Ser. A, Vol 239, No 809,
1944, p.392 }
\vskip5ex}


\begin{abstract}
Specializations of Schur functions are exploited to define and evaluate
the Schur functions $s_\lambda[\alpha X]$ and plethysms
$s_\lambda[\alpha s_\nu(X))]$ for any $\alpha$ - integer, real or complex.
Plethysms are then used to define pairs of mutually inverse infinite series 
of Schur functions, $M_\pi$ and $L_\pi$, specified by arbitrary partitions
$\pi$. These are used in turn to define and provide generating functions for
formal characters, $s_\lambda^{(\pi)}$, of certain groups $H_\pi$, 
thereby extending known results for orthogonal and symplectic group characters.
Each of these formal characters is then given a vertex operator realization,
first in terms of the series $M=M_{(0)}$ and various $L_\sigma^\perp$ dual to
$L_\sigma$, and then more explicitly in exponential form. Finally the
replicated form of such vertex operators are written down.
\end{abstract}

\maketitle


\section{Introduction}
\label{section:Introduction}

The aim here is to exploit the Hopf algebra structure of the ring
$\Lambda(X)$ of symmetric functions of the independent variables
$(x_1,x_2,\ldots)$, finite or countably infinite in number, that constitute 
the alphabet $X$. An emphasis will be placed on the interconnections
between the various products and coproducts that apply to the Schur
functions $s_\lambda(X)$ that form an integral basis of $\Lambda(X)$. These
allow us to define certain replicated, rational or scaled plethysms that
involve an argument $\alpha$ in $\openN$, $\openQ$ or $\openR$, respectively,
or even to $\openC$ or a sequence of such parameters.

The first key result is that, for any alphabet $X=(x_1,x_2,\ldots\,)$
and parameter $\alpha$, and any partitions $\lambda$ and $\nu$, we have
\begin{align}\label{result1}
 s_\lambda[\alpha s_\nu(X)] 
 &= \sum_{\mu,\rho}\, g_{\rho,\mu}^\lambda\, \dim_\mu(\alpha)\, 
    s_\mu[s_\nu(X)]\,, 
\end{align}
where the coefficients $g_{\rho,\mu}^\lambda$ are Kronecker coefficients
associated with products of characters of the symmetric group $S_m$ with
$m=|\lambda|$, the weight of the partition $\lambda$, while 
$\dim_\mu(\alpha)$ is the polynomial in $\alpha$ that gives the dimension of
the irreducible representation of $GL(n)$ specified by the partition $\mu$ 
evaluated at $n=\alpha$. The map $\dim : \Lambda(X) \longrightarrow R$
is an algebra homomorphism for any target ring $R$.

Following some notational preliminaries in Section~\ref{section:Notation}, this
result is obtained in Section~\ref{section:ParametrizedPlethysms} through the
use of one of the specializations introduced in
Section~\ref{section:Specializations}. 
Section~\ref{section:ParametrizedPlethysms} also contains some examples of
replicated plethysms and the computer benchmarking of their calculation, 
showing that the formula (\ref{result1}) is very efficient. 
The relevant algorithm is relegated to Appendix~\ref{appendix:code} in the form
of appropriate computer pseudo code.
In the special
case $\nu=(1)$ for which $s_\nu(X)=X$, the above plethysms coincide both with 
the replicated plethysms of Jarvis and Yung~\cite{jarvis:yung:1992a}
and, setting $\alpha=q$, with the $q$-analogues of Schur functions introduced
by Brenti~\cite{brenti:2000a}. Section~\ref{section:ParametrizedPlethysms}
includes an account of their orthogonality properties as given by
Baker~\cite{baker:1994a} and Brenti~\cite{brenti:2000a} but obtained here by
exploiting the Schur-Hall scalar product for the ring $\Lambda(X)$.

The next result realizes Littlewood's hope that an algebraic treatment of
character theory greatly generalizes the scope of the classical approach so
as to encompass cases which are very difficult to treat by analytical methods.
This same scalar product, in the form of the Cauchy identity, is then exploited
in Section~\ref{section:SeriesPlethysms} to derive the character generating
function
\begin{align}
L_\pi(Z)\, M(XZ) 
  &= \sum_\lambda\, s_\lambda^{(\pi)}(X)\, s_\lambda(Z)
\end{align}
for formal characters $s_\lambda^{(\pi)}(X)$ of $H_\pi$, each specified by 
a partition $\lambda$, where $H_\pi$ is the subgroup of the general
linear group preserving an invariant form of symmetry $\pi$, as introduced
elsewhere~\cite{fauser:jarvis:king:wybourne:2006a}. Here 
$X=(x_1,x_2,\ldots\,)$ is to be evaluated at the sequence of eigenvalues
of group elements of $H_\pi$.
The notation is such that 
$M(XZ)=\prod_{i,j\geq1}(1-x_iz_j)^{-1}$, while $L_\pi(Z)=L[s_\pi(Z)]$ is an
infinite Schur function series plethysm, with $L(Y)=\prod_{k\geq1}(1-y_k)$ for
all $Y$, including the case for which the elements $y_k$ of $Y$ are the
monomials of $s_\pi(Z)$. In this case, for an alphabet $Z$ of cardinality $l$, the cardinality
of $Y$ is exactly $\dim_\pi(l)$.

By exploiting the same series $L$, its inverse $M$ and its dual $L^\perp$ 
(i.e. its adjoint with respect to the Schur-Hall scalar product), together with the
Hopf algebra structure of $\Lambda(X)$, a vertex operator realization of the
characters $s_\lambda{\!\!}^{(\pi)}(X)$ is derived in
Section~\ref{section:VertexOperators}. This takes the form of another key
result, namely,
\begin{align}
s_\lambda^{(\pi)} 
  &= [Z^\lambda\,]\ V^\pi(z_1)\,V^\pi(z_2)\,\cdots\,V^\pi(z_l)\cdot 1\,,
\end{align}
where the vertex operators are given by
\begin{align}
V^\pi(z) 
  &= (1-z^p\,\delta_{\pi,(p)})\, M(z)\ L^\perp(z^{-1})\  
     \prod_{k=1}^{p-1}\, L^\perp_{\pi/(k)}(z^k) \,.
\end{align}

Then by means of exponential expressions for both $M$ and $L^\perp$, these
vertex operators are explicitly constructed in exponential form for all 
partitions $\pi$ of weight $|\pi|\leq 3$. The case $\pi=(2,1)$ is also given
an alternative normal ordering derivation in Section~\ref{section:VertexOperators}.
The explicit evaluation of $L^\perp_\pi(w)(M(Z))$ is undertaken in 
Appendix~\ref{appendix:LperpM}, with the result expressed, perhaps somewhat surprisingly,
in terms of semistandard Young tableaux of shape specified  by the partition
$\pi$. 

In Section~\ref{section:ReplicatedVertexOperators} it is pointed out rather
briefly that a wide class of replicated vertex operators may be obtained
very easily through the application of the replicated Schur functions of 
Section~\ref{section:ParametrizedPlethysms} to parametrized versions
of the vertex operators of Section~\ref{section:VertexOperators}.

Finally, Section~\ref{section:Conclusion} consists of a few concluding remarks.

\medskip
\section{Notational preliminaries}\label{section:Notation}
\setcounter{equation}{0}

\subsection{Partitions and Young diagrams}\label{subsection:partitions}
Our notation follows in large part that of Macdonald~\cite{macdonald:1995a}.  
Partitions are specified by lower case Greek letters. 
If $\lambda$ is a partition of $n$ we write $\lambda\vdash n$, and 
$\lambda=(\lambda_1,\lambda_2,\ldots,\lambda_n)$ is a sequence of 
non-negative integers $\lambda_i$ for $i=1,2,\ldots,n$
such that $\lambda_1\geq\lambda_2\geq\cdots\geq\lambda_n\geq0$, 
with $\lambda_1+\lambda_2+\cdots+\lambda_n=n$. The partition
$\lambda$ is said to be of weight $|\lambda|=n$ and length
$\ell(\lambda)$, where $\lambda_i>0$ for all $i\leq\ell(\lambda)$ 
and $\lambda_i=0$ for all $i>\ell(\lambda)$. In specifying $\lambda$
the trailing zeros, that is those parts $\lambda_i=0$, are often 
omitted, while repeated parts are sometimes written in exponent form
$\lambda=(\cdots,2^{m_2},1^{m_1})$ where $\lambda$ contains $m_i$
parts equal to $i$ for $i=1,2,\ldots$. For each such partition,
$n(\lambda)=\sum_{i=1}^n (i-1)\lambda_i$ and 
$z_\lambda=\prod_{i\geq1} i^{m_i}\, m_i!$.

Each partition $\lambda$ of weight $|\lambda|$ and length $\ell(\lambda)$
defines a Young or Ferrers diagram, $F^\lambda$, consisting of $|\lambda|$ 
boxes or nodes arranged in $\ell(\lambda)$ left-adjusted rows of lengths 
from top to bottom $\lambda_1,\lambda_2,\ldots,\lambda_{\ell(\lambda)}$ 
(in the English convention). The partition $\lambda'$, conjugate to 
$\lambda$, is the partition specifying the column lengths of $F^\lambda$ read
from left to right. The box $(i,j)\in F^\lambda$ in 
the $i$th row and $j$th column is said to have content 
$c(i,j)=j-i$ and hook length $h(i,j)=\lambda_i+\lambda'_j-i-j+1$. 

By way of illustration, if $\lambda=(4,2,2,1,0,0,0,0,0)=(4,2,2,1)=(4,2^2,1)$
then $|\lambda|=9$, $\ell(\lambda)=4$, $\lambda'=(4,3,1^2)$,
\begin{align}
F^\lambda
  &=F^{(4,2^2,1)} =\ \raisebox{-0.8cm}{\yng(4,2,2,1)}\ \ \hbox{and}\ \
    F^{\lambda'}=F^{(4,3,1^2)} =\ \raisebox{-0.8cm}{\yng(4,3,1,1)} \ \ .
\end{align}
The content and hook lengths of $F^\lambda$ are specified by
\begin{align}
  \raisebox{-0.8cm}{\young(0123,\10,\2\1,\3)}\ \qquad &\hbox{and}\qquad \ 
  \raisebox{-0.8cm}{\young(7521,42,31,1)}\ \ ,
\end{align}
where $\ov{m}=-m$ for all $m$. In addition, $n(4,2^2,1) = 0\cdot4+1\cdot2
+2\cdot2+3\cdot1=9$ and $z_{(4,2^2,1)} = 4\cdot2^2\cdot1\cdot1!\cdot2!\cdot1!
=32$.

\subsection{The ring $\Lambda(X)$ and Schur functions}
There exist various bases of $\Lambda(X)$ as described 
in~\cite{macdonald:1995a}: the monomial symmetric functions 
$\{m_\lambda\}_\lambda$, the complete symmetric functions 
$\{h_\lambda\}_\lambda$, the elementary symmetric functions 
$\{e_\lambda\}_\lambda$, the power sum symmetric functions
$\{p_\lambda\}_\lambda$ and the Schur symmetric functions 
$\{s_\lambda\}_\lambda$. Three of these bases are multiplicative,
with $h_{\lambda}=h_{\lambda_1}h_{\lambda_2}\cdots h_{\lambda_n}$, 
$e_{\lambda}=e_{\lambda_1}e_{\lambda_2}\cdots e_{\lambda_n}$ and 
$p_{\lambda}=p_{\lambda_1}p_{\lambda_2}\cdots p_{\lambda_n}$.
Of the relationships between the various bases we just mention at this stage 
the transitions 
\begin{align}
p_\rho(X) 
  &=\sum_{\lambda\vdash n}\ \chi^\lambda_\rho\ s_\lambda(X)
    &&\hbox{and}&&
    s_\lambda(X) = \sum_{\rho\vdash n}\ z_\rho^{-1}\ \chi^\lambda_\rho\ p_\rho(X)\,,
\label{Eq-p-s}
\end{align}
where $\chi^\lambda_\rho$ is the character of the irreducible representation
of the symmetric groups $S_n$ specified by $\lambda$ in the conjugacy class
specified by $\rho$. These characters satisfy the orthogonality conditions
\begin{align}
\sum_{\rho\vdash n}\ z_\rho^{-1}\ \chi^\lambda_\rho\ \chi^\mu_\rho\ 
  &= \ \delta_{\lambda,\mu}
    &&\hbox{and}&&
   \sum_{\lambda\vdash n}\ z_\rho^{-1}\ \chi^\lambda_\rho\ \chi^\lambda_\sigma\ = \ \delta_{\rho,\sigma}\,.
\label{Eq-chi-orth}
\end{align}

The significance of the Schur function basis lies in the fact that with respect
to the usual Schur-Hall scalar product 
$\langle \cdot \,|\, \cdot \rangle_{\Lambda(X)}$ on $\Lambda(X)$ we have
\begin{align}
\langle s_\mu(X) \,|\, s_\nu(X) \rangle_{\Lambda(X)} 
  &= \delta_{\mu,\nu}\,. 
\label{Eq-scalar-prod-s}
\end{align}
From (\ref{Eq-p-s}) and (\ref{Eq-chi-orth}) it follows that
\begin{align}
\langle p_\rho(X) \,|\, p_\sigma(X) \rangle_{\Lambda(X)} 
  &= z_\rho \delta_{\rho,\sigma}\,. 
\label{Eq-scalar-prod-p}
\end{align}

In what follows we shall make considerable use of several infinite series 
of Schur functions. The most important of these are the mutually inverse
pair defined by
\begin{align}
 M(t;X) 
   &= \prod_{i\geq1} (1-t\,x_i)^{-1} 
    = \sum_{k\geq0}\ h_m(X)\,t^m\,;\label{Eq-M}\\
 L(t;X) 
   &= \prod_{i\geq1} (1-t\,x_i) 
    = \sum_{k\geq0}\ (-1)^m\, e_m(X)\,t^m\,, \label{Eq-L}
\end{align} 
where as Schur functions $h_m(X)=s_{(m)}(X)$ and $e_m(X)=s_{(1^m)}(X)$. 
It might be noted that in Macdonald's notation and $\lambda$-ring 
notation $M(t;X)=H(t)=\sigma_t(X)$ and $L(t;X)=E(-t)=\lambda_{-t}(X)$.
For convenience, in the case $t=1$ we write $M(1;X)=M(X)$ and $L(1;X)=L(X)$.

\subsection{Algebraic properties of $\Lambda(X)$}
The ring, $\Lambda(X)$, of symmetric functions over $X$ has a Hopf algebra
structure, and two further algebraic and two coalgebraic operations. For
notation and basic properties we refer for example 
to~\cite{fauser:jarvis:2003a,fauser:jarvis:king:wybourne:2006a} and 
references therein. For the moment, in the interest of typographical
simplicity, the symbol $X$ for the underlying alphabet is suppressed unless
specifically required. 

We indicate outer products on $\Lambda$ either by $m$, or with infix notation 
using juxtaposition. Inner products are denoted either by $\textsf{m}$ or as
infix by $\ip$, while plethysms (compositions) are denoted by $\circ$ or by
means of square brackets $[\phantom{\ldots}]$. The corresponding coproduct
maps are specified by $\Delta$ for the outer coproduct, $\delta$ for the inner
coproduct, and $\PlethDelta$ for the plethysm coproduct. In Sweedler notation 
the action of these coproducts is distinguished by means of different brackets, 
round, square and angular, around the Sweedler indices -- the so-called
Brouder-Schmitt convention. The coproduct coefficients themselves are obtained
from the products by duality using the Schur-Hall scalar product and the
self-duality of $\Lambda(X)$. For example, for all $A,B\in\Lambda(X)$:
\begin{align}
\outer(A\otimes B) &= AB \,;
&&&
\CO(A) &= A_{(1)}\otimes A_{(2)} \,;\nonumber \\
\inner(A\otimes B) &= A\ip B\,; 
&&&
\co(A) &= A_{[1]}\otimes A_{[2]} \,;\nonumber \\
A\circ B &= A[B] \,;
&&&
\PlethDelta(A) &= A_{<1>}\otimes A_{<2>}\,. \nonumber  
\end{align}

In terms of the Schur function basis $\{s_\lambda\}_{\lambda\vdash
n,n\in\mathbb{N}}$ the product and coproduct maps give
rise to the particular sets of coefficients specified as follows:
\begin{align}
s_\mu s_\nu &= \sum_\lambda c^\lambda_{\mu,\nu}s_\lambda \,;
&&&
\CO(s_\lambda) 
  &= s_{\lambda_{(1)}}\otimes s_{\lambda_{(2)}}
   =\sum_{\mu,\nu}c^\lambda_{\mu,\nu} s_\mu\otimes s_\nu\,;
   \nonumber \\
s_\mu \ip s_\nu &= \sum_\lambda g^\lambda_{\mu,\nu}s_\lambda \,;
&&&
\co(s_\lambda) 
  &= s_{\lambda_{[1]}}\otimes s_{\lambda_{[2]}}
   =\sum_{\mu,\nu}g^\lambda_{\mu,\nu} s_\mu\otimes s_\nu \,;
   \nonumber \\
s_\mu[s_\nu] &= \sum_\lambda p^\lambda_{\mu,\nu}s_\lambda \,;
&&&
\PlethDelta(s_\lambda) 
  &= s_{\lambda_{\la 1\ra}}\otimes s_{\lambda_{\la 2\ra}}
   =\sum_{\mu,\nu}p^\lambda_{\mu,\nu} s_\mu\otimes s_\nu \,.
   \nonumber
\end{align}
Here the $c^\lambda_{\mu,\nu}$ are Littlewood-Richardson coefficients,
the $g^\lambda_{\mu,\nu}$ are Kronecker coefficients and the
$p^\lambda_{\mu,\nu}$ are plethysm coefficients. All these coefficients 
are non-negative integers. The Littlewood-Richardson coefficients
can be obtained, for example, by means of the Littlewood-Richardson 
rule~\cite{littlewood:richardson:1934a,littlewood:1950a} or the hive 
model~\cite{buch:2000a}. The Kronecker coefficients may determined directly
from characters of the symmetric group or by exploiting the Jacobi-Trudi
identity and the Littlewood-Richardson rule~\cite{robinson:1961a}, while
plethysm coefficients have been the subject of a variety methods of
calculation~\cite{littlewood:1950b,robinson:1961a,chen:garsia:remmel:1984a}.
Note that the above sums are finite, since
\begin{align}
c^\lambda_{\mu,\nu} &\ge0 
  \qquad\textrm{iff}\quad \vert\lambda\vert=\vert\mu\vert +\vert\nu\vert\,;
  \nonumber \\
g^\lambda_{\mu,\nu} &\ge0 
  \qquad\textrm{iff}\quad \vert\lambda\vert=\vert\mu\vert =\vert\nu\vert\,;
  \nonumber \\
p^\lambda_{\mu,\nu} &\ge0 
  \qquad\textrm{iff}\quad \vert\lambda\vert=\vert\mu\vert\,\vert\nu\vert\,.
  \nonumber
\end{align}

The Schur-Hall scalar product may be used to define skew Schur functions
$s_{\lambda/\mu}$ through the identities
\begin{align}
c^\lambda_{\mu,\nu}
  &= \la\, s_\mu\, s_\nu \,|\, s_\lambda\,\ra = \la\, s_\nu\,|\, 
        s^\perp_\mu (s_\lambda)\, \ra
   = \la \, s_\nu\,|\, s_{\lambda/\mu}\, \ra\,,
\label{Eq-def-skew}
\end{align}
so that 
\begin{align}
s_{\lambda/\mu} 
  &=  \sum_\nu\ c^\lambda_{\mu,\nu}\ s_\nu\,.
\label{Eq-skew}
\end{align}    

Within the outer product Hopf algebra we have a unit $\Id$, 
a counit $\varepsilon$ and an antipode $\antip$ such that\footnote{%
Macdonald uses the involution $\omega$ which differs from the antipode
by a sign factor: $\antip(s_\lambda)=(-1)^{\ell(\lambda)}\omega(s_\lambda)$. 
It is, however, convenient to employ the antipode if Hopf algebra structures
are in use.}:
\begin{align}
  \Id(1)
    &=s_0\,;\qquad \varepsilon(s_\lambda)
     =\delta_{\lambda,(0)}\,; \qquad \antip(s_\lambda) 
     = (-1)^{|\lambda|} s_{\lambda'}\,.
\label{Eq-antip}
\end{align} 

\subsection{The Cauchy kernel}
It is often convenient to represent an alphabet in an additive manner
$X=x_1+x_2+\cdots$, as itself an element of the ring $\Lambda(X)$ in the
sense that
\begin{align}
  X
    &= x_1+x_2+\cdots\ 
     = h_1(X) = e_1(X) = p_1(X) = s_{(1)}(X)\,. \nonumber
\end{align}
As elements of $\Lambda(X)\otimes\Lambda(Y)$ we have
\begin{align}
  X\!\!+\!\!Y
    &=x_1+x_2+\cdots+y_1+y_2+\cdots 
  \nonumber \\
  XY
    &=(x_1+x_2+\cdots)(y_1+y_2+\cdots)
     =(x_1y_1+x_1y_2+\cdots+x_2y_1+x_2y_t+\cdots)\,. 
  \nonumber
\end{align}

With this notation, the outer coproduct gives
\begin{align}
  \Delta(M)
    &=M_{(1)}\otimes M_{(2)} = M\otimes M
    &&&
      M(X\!\!+\!\!Y)
    &= \prod_i \frac{1}{1-x_i}\,\prod_j\,\frac{1}{1-y_j}\,;
    \nonumber \\
  \Delta(L)
    &=L_{(1)}\otimes L_{(2)} = L\otimes L
    &&&
      L(X\!\!+\!\!Y)
    &= \prod_i (1-x_i)\, \prod_j\,(1-y_j) \,,
    \nonumber
\end{align}  
so that
\begin{align}
M(X\!\!+\!\!Y)
  &=M(X)\,M(Y)\quad\hbox{and}\quad L(X\!\!+\!\!Y)=L(X)\,L(Y)\,.
\label{Eq-Delta-ML}  
\end{align}
For the inner coproduct: 
\begin{align}
  \delta(M)
     &=M_{[1]}\otimes M_{[2]}
     &&&  
       M(XY)
     &= \prod_{i,j} \frac{1}{1-x_iy_j}\,; 
   \nonumber \\
  \delta(L)
     &=L_{[1]}\otimes L_{[2]}
     &&&
       L(XY)
     &= \prod_{i,j} (1-x_iy_j) \,.
     \nonumber
\end{align}
The expansions of the products on the right hand sides of these expressions is
effected remarkably easily by evaluating the inner coproducts on the left: 
\begin{align}
\delta(M) 
  &= \sum_{k\geq0}\, \delta(h_k) 
   = \sum_{k\geq0} \sum_{\lambda\vdash k}\ s_\lambda \otimes s_\lambda\,;
  \nonumber \\
\delta(L) 
  &= \sum_{k\geq0}\,(-1)^k\,\delta(e_k) 
   = \sum_{k\geq0}\, (-1)^k\, \sum_{\lambda\vdash k}\ 
        s_\lambda\otimes s_{\lambda'}\,. 
  \nonumber 
\end{align}
This gives immediately the well known Cauchy and Cauchy-Binet formulae:
\begin{align}
  M(XY) 
    &= \prod_{i,j}\ \frac{1}{1-x_i\,y_j} 
     = \sum_\lambda\ s_\lambda(X)\, s_\lambda(Y)\,;
\label{Eq-Cauchy} 
    \\
  L(XY) 
    &= \prod_{i,j}\ (1-x_i\,y_j) 
     = \sum_\lambda\ (-1)^{|\lambda|} s_\lambda(X)\,s_{\lambda'}(Y) \,.
\label{Eq-dual-Cauchy}
\end{align}

That the Cauchy kernel, $M(XY)$, is a dual version of the Schur-Hall scalar
product can be seen by noting that
\begin{align}
s_\mu(X)\ M(XY) 
  &= \sum_\lambda\, \sum_\nu\, c^\nu_{\mu,\lambda}\ s_\nu(X)\, s_\lambda(Y)
  \nonumber \\ 
  &= \sum_\nu\, s_\nu(X)\, s_{\nu/\mu}(Y)
   = s^\perp_\mu(Y)\,(M(XY)). 
\label{Eq-sM-sperpM}
\end{align}
More generally, for any $F(X)\in \Lambda(X)$ with dual $F^\perp(X)$, by 
linearly extending the above result we have
\begin{align}
      F(X)\ M(XY) &= F^\perp(Y)\,(M(XY)). 
\label{Eq-FM-FperpM}
\end{align}
This is an identity that will be encountered and exploited a number 
of times in later sections. 
 
\subsection{Plethysms}
Plethysms are defined as compositions whereby for any $A,B\in\Lambda(X)$ the 
plethysm $A[B]$ is $A$ evaluated over an alphabet $Y$ whose letters are the 
monomials of $B(X)$, with each letter repeated as many times as the
multiplicity of the corresponding monomial. Thus the Schur function plethysm
is defined by
\begin{align}
  s_\lambda[s_\mu](X) 
    &=s_\lambda(Y)  &\hbox{where}&& Y=s_\mu(X)\,.
\label{Eq-sf-pleth}    
\end{align}  

For all $A,B,C\in \Lambda(X)$ we have the following rules, due to 
Littlewood~\cite{littlewood:1950a}, for manipulating plethysms:
\begin{align}
\label{littlewood:plethysm}
 (A+B)[C] &= A[C] + B[C]\,; &&&
  A[B+C]  &= A_{(1)}[B]A_{(2)}[C]\,;
 \nn
 (AB)[C]  &= A[C]B[C]\,;  &&&
 A[BC]    &= A_{[1]}[B]A_{[2]}[C]\,;
 \nn
 A[B[C]]  &= (A[B])[C]\,.
\end{align}
To these we can add, see~\cite{fauser:jarvis:king:wybourne:2006a}:
\begin{align}
\label{plethysm:antip:coprods}
 A[-B] &= (\antip(A))[B]\,; &&&
 A[\antip(B)]&= \antip(A[B])\,; 
\nonumber \\
 A[\CO(B)] &= \CO(A[B])\,; &&& 
 A[\co(B)] &= \co(A[B])\,, 
\end{align}
and the plethysm of a tensor product:
\begin{equation}
\label{plethysm:tensorprod}
A[B\otimes C]=A_{[1]}[B]\otimes A_{[2]}[C]\,. 
\end{equation} 

These rules enable us to evaluate plethysms not only of outer and inner
products but also of outer and inner coproducts. 
\medskip

\section{Specializations}\label{section:Specializations}
\setcounter{equation}{0}

\subsection{Definition of specializations}
Before dealing with the plethysms of interest here, it is appropriate to define
certain specializations. We will do this in some generality so as to be able to
use the technique of specialization in a rather broad context, see 
also~\cite[Sect.1.12]{macdonald:1999a}.

\mybenv{Definition}
A \textbf{specialization} $\phi$ is an algebra homomorphism (and thus a
1-cocycle for the outer Hopf algebra) from the Hopf algebra of the ring 
$\Lambda(X)$ of symmetric functions to another ring $R$, where $R$ may be any 
one of $\mathbb{N}, \mathbb{Z}, \mathbb{Q},
\mathbb{R}, \mathbb{C}, \mathbb{Z}[t], \mathbb{Z}[[q]],, \mathbb{Z}[t][[q]],
\ldots $. For any
$A,B\in \Lambda(X)$ it is required that we have
\begin{align}
\phi 
  &: \Lambda \rightarrow R 
  &&\hbox{with}&& 
  \phi(AB) = \phi(A)\phi(B) \,. \label{Eq-phi-spec}
\end{align}
\myeenv

Specializations $\phi$ may be defined either through a map, also denoted by
$\phi$, on the letters $x_i$ of the underlying alphabet $X=(x_1,x_2,\ldots)
= x_1+x_2+\cdots$, or through maps on the generators of $\Lambda(X)$ such as 
$h_n(X)$, $e_n(X)$ or $p_n(X)$.
  \subsection{Fundamental specialization}
  We denote by $\epsilon^1$ the map $\epsilon^1\circ A(X) = A(\epsilon^1(X))$
  where $\epsilon^1(X)=(1,0,\ldots\,)=1$, that is to say
  \begin{align}
    \epsilon^1(x_i) 
     &=\left\{
       \begin{array}{ll}
         1 & \textrm{if}~~i=1\,; \\
         0 & \textrm{otherwise}\,.
       \end{array} 
       \right. 
  \label{Eq-eps-1-x}
  \end{align}
  This fundamental specialization evaluates on Schur functions as the dimension
  formula for $GL(1)$ in the sense that 
  \begin{align}
    \epsilon^1(s_\lambda(X))
      &=s_\lambda(1,0,\ldots\,) 
       = \dim\,V_{GL(1)}^\lambda 
       = \dim_\lambda(1)\,,
  \label{Eq-eps-1-s1}
  \end{align}
  where $\dim\,V^\lambda_{GL(1)}$ is the dimension of the irreducible representation $V^\lambda_{GL(1)}$
  of $GL(1)$ having highest weight $\lambda$. This specialization is such that
  \begin{align}
  \epsilon^1(s_\lambda(X)) 
    &=\left\{
      \begin{array}{ll}
        1 & \textrm{if}~\lambda=(m)~\textrm{for any}~m\geq0; \\
        0 & \textrm{otherwise}.
      \end{array}
      \right.
  \label{Eq-eps-1-s2}
  \end{align}
  \subsection{$t$-specialization}
  We generalize the fundamental specialization along the following lines. For
  all $t\in\mathbb{N}$ we denote by
  $\epsilon^t\circ A(X)=A(\epsilon^t(X))$ with $\epsilon^t(X) = 
  (1,\ldots,1,0\,\ldots\,)=1+1+\cdots+1$ with $t$ occurrences of the $1$s. In
  the sequence notation we thus have $\epsilon^t\circ A(X)=A(1^t)$ while in
  the additive (ring) notation we have $\epsilon^t\circ A(X)=A[t]$. In both
  cases we make this precise as
  \begin{align}
  \epsilon^t(x_i)
    &=\left\{
      \begin{array}{ll}
         1 & \textrm{if}~1\le i \le t\,; \\
         0 & \textrm{otherwise}\,.
      \end{array} 
      \right.
  \label{Eq-eps-t-x}
  \end{align}
  For all $t\in \mathbb{N}$ the $t$-specialization of a Schur function can be
  interpreted by means of a $GL(t)$ dimension formula, that is
  $\epsilon^t(s_\lambda(X))= s_\lambda(1^t)=s_\lambda(1,1,\ldots,1) = 
  \dim V^\lambda_{GL(t)}=\dim_\lambda(t)$. However, the dimension formula for
  $GL(t)$ is polynomial in $t$:
  \begin{align}
     \dim_\lambda(t) 
       &= \prod_{(i,j)\in F^\lambda} \frac{t+c(i,j)}{h(i,j)}\,,
  \label{Eq-dim-t}
  \end{align}
  and hence can be generalized by analytic continuation to rational, real or
  even complex $t$.
  \subsection{Principal $(q;n)$-specialization}
  A further important specialization is given by the map 
  $\epsilon_{q;n}^1 \circ A(X) = A(1,q,q^2,\ldots,q^{n-1},0,\ldots\,) 
  = A[\frac{1-q^n}{1-q}]$ or
  \begin{align}
  \epsilon_{q;n}^1(x_i) 
    &=\left\{
      \begin{array}{ll}
        q^{i-1} &\textrm{for}~1\le i \le n; \\
        0 & \textrm{otherwise}.
      \end{array}
      \right.
  \label{Eq-eps-1qn-x}
  \end{align}
  
  In the case of Schur functions, with the notation described earlier, we
  have~\cite[p.44]{macdonald:1995a}
  \begin{align}
  \epsilon_{q;n}^1\circ s_\lambda(X)
    &= s_\lambda(1,q,\ldots,q^{n-1})
     = q^{n(\lambda)}\prod_{(i,j)\in F^\lambda} 
        \frac{1-q^{n+c(i,j)}}{1-q^{h(i,j)}}\,.
  \label{Eq-eps-1qn-s}
  \end{align}
  In the special case of $\lambda=(1^m)$ this takes the form
  \begin{align}
  \epsilon_{q;n}^1\circ s_{(1^m)}(X)
    &= \epsilon_{q;n}^1\circ e_m(X)
     = q^{m(m-1)/2}  
    \left[\!\!\!\begin{array}{c}
                      n\\ m
                \end{array}\!\!\!\right]_q\,, 
    \nonumber
  \end{align}
  where the $q$-binomial coefficient is given by
  \begin{align}
  \left[\!\!\!\begin{array}{c}n\\ m\\\end{array}\!\!\!\right]_q 
    &= \frac{(1-q^n)(1-q^{n-1})\cdots(1-q^{n-m+1})}{(1-q)(1-q^2)\cdots(1-q^m)}
       \,.  
    \nonumber
  \end{align}
  \subsection{Three parameter specialization}
  Note that for $q\rightarrow 1$ we recover the $t$-specialization from the
  principal $(q;t)$-specialization. However, we keep these two specializations
  apart so as to have the opportunity to employ a combination of both. This
  will be denoted by
  \begin{align}
  \epsilon^t_{q;n}\circ A(X) 
    &= A(1,\ldots,1,q,\ldots,q,\ldots,q^{n-1},\ldots,q^{n-1},0,\ldots)
     = A\left[t\,\frac{1-q^n}{1-q}\right]\,,
  \label{Eq-eps-tqn-A}
  \end{align}
  with $t$ repetitions of each distinct power of $q$, while $t=1+1+\cdots+1$ 
  with $t$ repetitions of $1$.
\medskip  

\section{Parameterized plethysms}\label{section:ParametrizedPlethysms}
\setcounter{equation}{0}

The idea now is to exploit the above findings to see if we can derive a general
formula for the plethysm $s_\lambda[\alpha\, s_\nu]$ for any $\alpha$: integer,
rational or complex. 

\subsection{Replicated Schur functions as plethysms}
First we deal with the case $\alpha=t\in\mathbb{N}$. In this case we may use
the iterated outer coproduct identity
\begin{align}
\Delta^{(t-1)} s_\lambda 
  &= (\Id\otimes \Delta^{(t-2)}) \Delta s_\lambda
   = (\Id\otimes \Delta^{(t-2)}) s_{\lambda_{(1)}}\otimes s_{\lambda_{(2)}}
   \nonumber \\
  &= s_{\lambda_{(1)}}\otimes (\Delta^{(t-2)}s_{\lambda_{(2)}})
   = \cdots 
   = s_{\lambda_{(1)}}\otimes s_{\lambda_{(2)}}\otimes \cdots \otimes 
      s_{\lambda_{(t)}}\,,
\label{Eq-Delta-t-1}
\end{align}
with $\Delta^{(2)}=\Delta$, $\Delta^{(1)}=\Id$, $\Delta^{(0)}=\epsilon^0$ and 
some relabelling has been applied to the iterated outer product Sweedler 
indices. Then using outer product multiplication $t-1$ times one finds
\begin{align}
s_\lambda[t\,s_\nu(X)] 
  & = s_\lambda[s_\nu(X)+s_\nu(X)+\cdots+s_\nu(X)]
  \nonumber \\
  & = s_{\lambda_{(1)}}[s_\nu(X)]\ s_{\lambda_{(2)}}[s_\nu(X)]\ \cdots 
     s_{\lambda_{(t)}}[s_\nu(X)]\,.
\label{Eq-tpleth-outer}
\end{align}

\mybenv{Example} 
\begin{align}
s_{(2)}[2 s_{(2)}] 
  &= s_{(2)}[s_{(2)}+s_{(2)}] = s_{(2)}[s_{(2)}]+ s_{(1)}[s_{(2)}]\ s_{(1)}[s_{(2)}]+ s_{(2)}[s_{(2)}]
  \nonumber \\
  &= (s_{(4)}+s_{(2,2)})+(s_{(4)}+s_{(3,1)}+s_{(2,2)})+(s_{(4)}+s_{(2,2)})
  \nonumber \\
  &=  3 s_{(4)} +s_{(3,1)} + 3 s_{(2,2)}\,;
  \nn
s_{(1,1)}[2 s_{(2)}] 
  &= s_{(1,1)}[s_{(2)}+s_{(2)}] = s_{(1,1)}[s_{(2)}]+ s_{(1)}[s_{(2)}]\ s_{(1)}[s_{(2)}]+ s_{(1,1)}[s_{(2)}]
  \nonumber \\
  &= s_{(3,1)}+(s_{(4)}+s_{(3,1)}+s_{(2,2)})+s_{(3,1)}
  \nonumber \\
  &=  s_{(4)} + 3 s_{(3,1)} + s_{(2,2)} \,. \nonumber
\end{align}
\myeenv%

Alternatively, we may use the inner coproduct identity
$\delta s_\lambda=s_{\lambda_{[1]}}\otimes s_{\lambda_{[2]}}$ to obtain
\begin{align}
s_\lambda[t\,s_\nu(X)] 
  &= s_{\lambda_{[1]}}[t]\ s_{\lambda_{[2]}}[s_\nu(X)]
   = \sum_{\rho,\mu}\, g^\lambda_{\rho,\mu}\ s_\rho[t]\ s_\mu[s_\nu(X)]
  \nonumber \\
  &= \sum_{\rho,\mu}\, g^\lambda_{\rho,\mu}\ \dim_\rho(t)\ s_\mu[s_\nu](X)
   =\sum_\mu\ b^\lambda_\mu(t)\ s_\mu[s_\nu](X)\,,
\label{Eq-tpleth-inner}
\end{align}
where 
\begin{align}
     b^\lambda_\mu(t) &= \sum_{\rho}\, g^\lambda_{\rho,\mu}\ \dim_\rho(t)\,.
\label{Eq-bt}
\end{align}

\mybenv{Example} 
\begin{align}
s_{(2)}[2 s_{(2)}] 
  &= \dim_{(2)}(2)\ s_{(2)}[s_{(2)}]+ \dim_{(1,1)}(2)\ s_{(1,1)}[s_{(2)}]
  \nonumber \\
  &= 3\,(s_{(4)}+s_{(2,2)})+(s_{(4)}+1\,s_{(3,1)}
  \nonumber \\
  &=  3 s_{(4)} +s_{(3,1)} + 3 s_{(2,2)}\,;
  \nn
s_{(1,1)}[2 s_{(2)}] 
  &= \dim_{(1,1)}(2)\ s_{(2)}[s_{(2)}]+ \dim_{(2)}(2)\ s_{(1,1)}[s_{(2)}]\,;
  \nonumber \\
  &= 1,(s_{(4)}+s_{(2,2)})+3\,s_{(3,1)}
  \nonumber \\
  &=  s_{(4)} + 3 s_{(3,1)} + s_{(2,2)}\,, \nonumber
\end{align}
as before.
\myeenv
\bigskip

In the special case $\nu=(1)$, for which $s_\nu(X)=X$, (\ref{Eq-tpleth-inner}) 
gives
\begin{align}
 s_\lambda[t\,X] 
   &= \sum_{\rho,\mu}\, g^\lambda_{\rho,\mu}\ \dim_\rho(t)\ s_\mu(X)
    = \sum_\mu\ b^\lambda_\mu(t)\ s_\mu(X)\,.
\label{Eq-t-schur}
\end{align}

\samepage{
\mybenv{Example}
\vskip-3ex 
\begin{align}
s_{(2)}[2\,X] 
  &= \dim_{(2)}(2)\ s_{(2)}(X)+ \dim_{(1,1)}(2)\ s_{(1,1)}(X)
  \nonumber \\
  &= 3\,s_{(2)}(X)+ s_{(1,1)}(X)\,;
  \nonumber \\
  \nn
s_{(1,1)}[2\,X] 
  &= \dim_{(1,1)}(2)\ s_{(2)}(X)+ \dim_{(2)}(2)\ s_{(1,1)}(X)
  \nonumber \\
  &=  s_{(2)}(X) + 3 s_{(1,1)}(X)\,.  \nonumber
\end{align}
\vskip-3ex
\myeenv
}

\subsection{Benchmarking replicated plethysm calculations}
The above shows that we may use either iterated outer coproducts, or a single
inner coproduct augmented by a dimensionality formula, to evaluate replicated
plethysms. Although the above examples might suggest that these two methods
are comparable in complexity, this is far from being the case. The iteration
may be very tedious, with the second method much more efficient, at least for
sufficiently large $n$. Symbolic computations show a dramatic increase of
speed for even modestly large $n$ (greater than 10). The relevant algorithm is 
given in the form of computer pseudo code in Appendix~\ref{appendix:code}. 

We have investigated this process via the use of both Maple using the 
\texttt{SchurFkt} package~\cite{schurfkt:2003a} and the open source software
\texttt{SCHUR}~\cite{SCHUR}. In arbitrary time units we can compare the 
computation of the plethysms as shown in Table~\ref{table1}.
\begin{table}[h]
\caption{Timing of the iterated and directly evaluated plethysms
$s_{(3)}[n\cdot s_{(1,1)}]$ using \texttt{SchurFkt}. (Note that figures are
obscured by Maple's garbage collection and not as accurate as 
shown)\label{table1}}
\begin{tabular}{lll}
\hline\hline
multiplicity & recursive & direct \\
\hline
n=1 	& 0.01 &  0.02\\
n=10 	& 0.08 &  0.02\\
n=100 	& 0.89 &  0.01\\
n=1000 	& 7.32 &  0.01\\
n=10000 &  --- &  0.01\\
\hline
\end{tabular}
\end{table}
Both algorithms make use of Maple remember tables, so a plethysm
is never computed twice. It is clear that the second
method is $O(1)$ with respect to $n$, while the first one increases
rapidly. Maple fails to do the iteration for $n=10000$. Very similar 
results can be obtained by using \texttt{SCHUR}, but the inner coproduct
case has to be carried out in two stages in order to insert the 
dimensionality factors appropriately.
\medskip

\subsection{$\alpha$-plethysms and $\alpha$-Schur functions}
Since the coefficients $b^\lambda_\mu(t)$ are polynomials in $t$, the 
formulae (\ref{Eq-tpleth-inner}) and (\ref{Eq-t-schur}) may be extended so as
to define $\alpha$-plethysms and $\alpha$-Schur functions by means of the
formulae
\begin{align}
 s_\lambda[\alpha\,s_\nu(X)] 
  &  = \sum_\mu\ b^\lambda_\mu(\alpha)\ s_\mu[s_\nu](X)\,
\label{Eq-alpha-pleth}
\end{align}
and
\begin{align}
 s_\lambda[\alpha\,X]
  & = \sum_\mu\ b^\lambda_\mu(\alpha)\ s_\mu(X)\,,
\label{Eq-alpha-schur}
\end{align}
where
\begin{align}
 b^\lambda_\mu(\alpha) 
   &= \sum_{\rho}\, g^\lambda_{\rho,\mu}\ \dim_\rho(\alpha)\,.
\label{Eq-alpha-b}
\end{align}

The symmetric functions $s_\lambda[\alpha X]$ are precisely those that were
introduced and studied by Baker~\cite{baker:1994a} as replicated Schur
functions, and independently by Brenti~\cite{brenti:2000a} as $q$-analogues of Schur
functions. Our notation is such that $s_\lambda[\alpha\,X]$ is identical to
Baker's $s_\lambda(x^{(\alpha)})$ and Brenti's $s_\lambda[x]_q$  
under the identifications $x=X$ and $q=\alpha$. 

The case of replicated and $\alpha$-power sum functions is even easier.
Since $p_n(X)=x_1^n+x_2^n+\cdots$, it follows immediately that for any
$t,n\in\mathbb{N}$ we have
\begin{align}
 p_n(t\,X)
   &= p_n(X+X+\cdots+X)= t\,(x_1^n+x_2^n+\cdots\,)= t\ p_n(X)\,,
\label{Eq-pn-t}
\end{align}
so that, replacing $X$ by $Y=p_\mu(X)$, we have
\begin{align}
 p_n(t\,p_\mu(X))
   &= t\ p_n[p_\mu(X)]\,.
\label{Eq-pn-tpleth}
\end{align}
The multiplicative nature of 
$p_\lambda=p_{\lambda_1}\,p_{\lambda_2}\,\cdots\,p_{\lambda_{\ell(\lambda)}}$,
where $\ell(\lambda)$ is the number of non-zero parts of $\lambda$,
is then such that:
\begin{align}
p_\lambda[t\,X]
  &= t^{\ell(\lambda)}\ p_\lambda(X)\,;\label{Eq-plambda-t}\\
p_\lambda[t\,p_\mu(X)]
  &= t^{\ell(\lambda)}\ p_\lambda[p_\mu(X)]\,.\label{Eq-plambda-tpleth}
\end{align}

Once again we are at liberty to extend the domain of $t$ to give,
as a matter of definition:
\begin{align}
p_\lambda[\alpha\,X]
  &= \alpha^{\ell(\lambda)}\ p_\lambda(X)\,;\label{Eq-plambda-alpha}\\
p_\lambda[\alpha\,p_\mu(X)]
  &= \alpha^{\ell(\lambda)}\ p_\lambda[p_\mu(X)]\,.\label{Eq-plambda-alpha-pleth}
\end{align}
The first of these is really the starting point in Brenti's development of
$q$-analogues of symmetric functions, and both Baker~\cite{baker:1994a} and 
Brenti~\cite{brenti:2000a} 
have pointed out that the Jack symmetric functions $J_{(n)}(X;\alpha^{-1})$ can
be expressed in the form
\begin{align}
  J_{(n)}(X;\alpha^{-1})
    &= \frac{n!}{\alpha^n}\ s_{(n)}[\alpha\,X]\,,
\label{Eq-Jack}
\end{align}
which specialize to zonal symmetric functions for $\alpha=2$.

\subsection{Orthogonality properties of $\alpha$-Schur functions}
\label{subsection:Orthogonality}
We may use the Schur-Hall scalar product
to extract from (\ref{Eq-alpha-schur}) the formula
\begin{align}
  b^\lambda_\mu(\alpha)
   &= \la\, s_\mu(X)\,,\, s_\lambda[\alpha\,X] \,\ra_{\Lambda(X)}
   \nonumber \\
   &= \sum_{\sigma,\tau}\ \chi_\sigma^\mu\ \chi_\tau^\lambda\ \la\, 
           p_\sigma(X)\,,\, p_\tau(\alpha\,X) \,\ra_{\Lambda(X)}
   \nonumber \\
   &= \sum_{\sigma,\tau}\ \chi_\sigma^\mu\ \chi_\tau^\lambda\ \alpha^{\ell(\tau)} 
       \la\, p_\sigma(X)\,,\, p_\tau(X)\, \ra_{\Lambda(X)}
   \nonumber \\
   &= \sum_{\sigma,\tau}\ \chi_\sigma^\mu\ \chi_\tau^\lambda\ \alpha^{\ell(\tau)}
       \ z_\tau^{-1}\ \delta_{\sigma,\tau}
    = \sum_\sigma\ z_\sigma^{-1}\ \chi_\sigma^\mu\ \chi_\sigma^\lambda\ 
       \alpha^{\ell(\sigma)}\,,
\label{Eq-b-chi} 
\end{align}
where use has been made of (\ref{Eq-p-s}) and (\ref{Eq-plambda-alpha}).

With this determination of the coefficients $b^\lambda_\mu(\alpha)$ we can establish the
following result due to Baker~\cite{baker:1994a} and Brenti~\cite{brenti:2000a}:

\mybenv{Theorem}\ \ 
For all non-zero $\alpha$
\begin{equation}
   \la\, s_\mu(\alpha\,X)\,,\, s_\lambda(\alpha^{-1}\,X) \,\ra_{\Lambda(X)} \ = \ \delta_{\mu,\lambda}\,.
   \label{Eq-alpha-schur-orth}
\end{equation}
\myeenv

\noindent{\bf Proof:}
\begin{align}
  & \la\, s_\mu(\alpha\,X)\,,\, s_\lambda(\beta\,X) \,\ra_{\Lambda(X)}
  \nonumber \\
  &= \sum_{\sigma,\tau}\ \chi_\sigma^\mu\ \chi_\tau^\lambda\ \la\, 
       p_\sigma(\alpha\,X)\,,\, p_\tau(\beta\,X) \,\ra_{\Lambda(X)}
  \nonumber \\
  &= \sum_{\sigma,\tau}\ \chi_\sigma^\mu\ \chi_\tau^\lambda\ 
      \alpha^{\ell(\sigma)} \beta^{\ell(\tau)}  \la\, p_\sigma(X)\,,\, 
      p_\tau(X)\, \ra 
  \nonumber \\
  &= \sum_{\sigma,\tau}\ \chi_\sigma^\mu\ \chi_\tau^\lambda\ 
      \alpha^{\ell(\sigma)} \beta^{\ell(\tau)}\ z_\tau^{-1}\ 
      \delta_{\sigma,\tau} 
   = \sum_\sigma\ z_\sigma^{-1}\ \chi_\sigma^\mu\ \chi_\sigma^\lambda\ 
      (\alpha\beta)^{\ell(\sigma)}\,.\nonumber
\end{align}
Hence, taking $\beta=1/\alpha$ we have the Baker-Brenti orthogonality condition
\samepage{
\begin{align}
   \la\, s_\mu(\alpha\,X)\,,\, s_\lambda(\alpha^{-1}\,X) \,\ra_{\Lambda(X)} 
   &= \sum_\sigma\ z_\sigma^{-1}\ \chi_\sigma^\mu\ \chi_\sigma^\lambda\  
    = \delta_{\mu,\lambda}\,.\nonumber
\end{align}
\qed}

Now consider the following technical result.
\mybenv{Lemma}
For any positive integer $n$ and any partitions $\nu$ and $\rho$ of the same
weight
\begin{align}
 n^{\ell(\rho)}\, \chi^\nu_\rho
   &= \sum_{\mu,\sigma,\tau,\ldots,\zeta}\ \chi^\mu_\rho\ 
    c^\mu_{\sigma,\tau,\ldots,\zeta}\  c^\nu_{\sigma,\tau,\ldots,\zeta}\,,
\label{Eq-tech-lemma}
\end{align}
where the sum is over $n+1$ partitions $\mu,\sigma,\tau,\ldots,\zeta$.
\myeenv

\noindent{\bf Proof:} Consider
\begin{align}
 p_\rho(X,Y,\ldots,Z)
   &= \sum_\mu\ \chi^\mu_\rho\ s_\mu(X,Y,\ldots,Z)
  \nonumber \\
   &= \sum_{\mu,\sigma,\tau,\ldots,\zeta}\ 
      \chi^\mu_\rho\ c^\mu_{\sigma,\tau\ldots,\zeta}\ s_\sigma(X)\ s_\tau(Y) 
      \cdots s_\zeta(Z)
  \nonumber
\end{align}
where the coproduct $\Delta$ has been applied $n-1$ times. If we now apply the
multiplication operator $n-1$ times, that is we set $X = Y = \cdots = Z$ and
take outer products, we obtain
\begin{align}
 p_\rho(X,X,\ldots,X)
   &= p_\rho(n\,X)\ 
    = n^{\ell(\rho)}\ p_\rho(X) 
    = n^{\ell(\rho)}\ \sum_\nu\ \chi^\nu_\rho \ s_\nu(X) 
  \nonumber \\
   &= \sum_{\mu,\sigma,\tau,\ldots,\zeta}\ \chi^\mu_\rho\ 
      c^\mu_{\sigma,\tau\ldots,\zeta}\ 
      c^\nu_{\sigma,\tau\ldots,\zeta}\ s_\nu(X) \,.
  \nonumber
\end{align}
Comparing the coefficients of $s_\nu(X)$ and using (\ref{Eq-tech-lemma}) 
proves the Lemma.
\qed

This Lemma allows us to prove the following more general orthogonality
theorem:
\mybenv{Theorem}~\cite{baker:1994a} For $n$ alphabets $X,Y,\ldots,Z$, let  
\begin{align}
  \tilde{s}_\lambda(X,Y,\ldots,Z)
    &= \sum_{\mu}\ b_{\lambda,\mu}(1/n)\ s_\mu(X,Y,\ldots,Z)\,.
\label{Eq-stilde}
\end{align}
Then
\begin{align}
 \la\, \tilde{s}_\lambda(X,Y,\ldots,Z)\,,\, 
 s_\nu(X,Y,\ldots,Z)\,
   \ra_{\Lambda(X)\otimes\Lambda(Y)\otimes\cdots\otimes\Lambda(Z)} 
   &= \delta_{\lambda,\nu}\,.
\label{Eq-stilde-orth}
\end{align}
\myeenv

\noindent{\bf Proof:}
\begin{align}
  \la\, \tilde{s}_\lambda(X,&Y,\ldots,Z)\,,\, s_\nu(X,Y,\ldots,Z)\,
      \ra_{\Lambda(X)\otimes\Lambda(Y)\otimes\cdots\otimes\Lambda(Z)} 
  \nonumber \\
  &= \sum_\mu\ b_{\lambda,\mu}(1/n)\
     \la\, s_\mu(X,Y,\ldots,Z)\,,\, s_\nu(X,Y,\ldots,Z)\,
        \ra_{\Lambda(X)\otimes\Lambda(Y)\otimes\cdots\otimes\Lambda(Z)} 
  \nonumber \\
  &= \sum_\mu\ b_{\lambda,\mu}(1/n)\
  \sum_{\sigma,\tau,\ldots,\zeta,\eta,\theta\ldots,\phi}\ 
    c^\mu_{\sigma,\tau,\ldots,\zeta}\ 
    c^\nu_{\eta,\theta\ldots,\phi}\
   \delta_{\sigma,\eta}\ \delta_{\tau,\theta}\ \cdots\ \delta_{\zeta,\phi}
  \nonumber \\
  &= \sum_{\mu,\rho}\ z_\rho^{-1}\ \chi^\lambda_\rho\ \chi^\mu_\rho\ 
     n^{-\ell(\rho)}\ 
     \sum_{\sigma,\tau,\ldots,\zeta}\ 
       c^\mu_{\sigma,\tau,\ldots,\zeta}\ 
       c^\nu_{\sigma,\tau,\ldots,\zeta}
  \nonumber \\ 
  &= \sum_\rho\ z_\rho^{-1}\ \chi^\lambda_\rho\ \chi^\nu_\rho 
  \nonumber \\
  &= \delta_{\lambda,\nu}\,.
\nonumber
\end{align}
\qed

\medskip
\section{Series plethysms and character generating functions}
\label{section:SeriesPlethysms}
\setcounter{equation}{0}

\subsection{Series defined by plethysms}
Given  
\begin{align}
  M(t;X)
    &= \prod_{i\geq1} (1-t\,x_i)^{-1}
     =\sum_{k=0}^\infty\ t^k\, s_{(k)}(X)\,,
\end{align} 
new Schur function series may be generated from $M(t;X)$ by means of the
plethysm operation, as explained 
elsewhere~\cite{fauser:jarvis:king:wybourne:2006a}. For each fixed partition
$\pi$ one merely replaces $X$ by $Y=s_\pi(X)$ to give
\begin{align}
 M_\pi(t;X)
   &= M(t;[s_\pi])(X)=M(t;s_\pi(X))
   \nonumber \\
   &= M(t;Y)
    =\prod_{j\geq1}\, (1-t\,y_j)^{-1}
    = \sum_{k=0}^\infty\ t^k\, s_{(k)}[s_\pi](X)\,,
\end{align}
where the product is taken over all monomials $y_j$ of $s_\pi(X)$. Similarly,
from
\begin{align}
 L(t;X)
   &= \prod_{i\geq1} (1-t\,x_i)
    = \sum_{k=0}^\infty (-1)^k t^k\, s_{(1^k)}(X) \,,
\end{align} 
\vfill
\newpage
one obtains
\begin{align}
 L_\pi(t;X)
   &= L(t;[s_\pi])(X)
    = L(t;s_\pi(X))
   \nonumber \\
   &= L(t;Y)
    = \prod_{j\geq1}\, (1-t\,y_j)
    = \sum_{k=0}^\infty (-1)^k t^k\, s_{(1^k)}[s_\pi](X)\,.
\end{align}

\subsection{Character generating functions and the $M_\pi$ and $L_\pi$ series} 
The Cauchy kernel $M(XZ)$ serves as a generating function for characters of
$GL(n)$ in the sense that
\begin{align}
 M(XZ)
  &= \prod_{i,j} (1-x_iz_j)^{-1} 
   = \sum_\lambda\ s_\lambda(X)\ s_\lambda(Z)\,,
\end{align} 
where $s_\lambda(X)$ is the character of the irreducible representation
$V^\lambda_{GL(n)}$ of highest weight $\lambda$ evaluated at group elements
whose eigenvalues are the elements of $X$. As pointed out earlier this implies,
and is implied by
\begin{align}
 s_\lambda(X) 
   &= [s_\lambda(Z)](\, M(XZ)\,) 
    = \la\, s_\lambda(Z)\,|\, M(XZ)\, \ra_{\Lambda(Z)} \,,
\end{align}
where $[s_\lambda](\, \cdots\,)$ denotes the coefficient of $s_\lambda(Z)$ in
$(\, \cdots\,)$. 

Now we are in a position to determine the analogous generating functions for 
certain formal characters, $s_\lambda^{(\pi)}(X)$, of subgroups $H_\pi(n)$ of
$GL(n)$ introduced elsewhere~\cite{fauser:jarvis:king:wybourne:2006a}
by exploiting the mutually inverse series $M_\pi=M[s_\pi]$ and 
$L_\pi=L[s_\pi]$. To be more precise, we let
\begin{align}
 s_\lambda^{(\pi)}(X)
  &= L^\perp_\pi(X)\,(s_\lambda(X)).
\label{Eq-def-slambda-pi}
\end{align}
The generating function for these characters may then be found as follows:
\begin{align}
 s_\lambda^{(\pi)}(X)
   &= L^\perp_\pi(X)\left(s_\lambda(X)\right) 
    = L^\perp_\pi(X)\left([s_\lambda(Z)]\, M(XZ)\right)
    \nonumber \\
   &= [s_\lambda(Z)]\ L^\perp_\pi(X)\,(M(XZ))\,.
\label{Eq-slambda-pi}
\end{align}
It then follows from (\ref{Eq-FM-FperpM}) that
\begin{align}
 s_\lambda^{(\pi)}(X)
   &= [s_\lambda(Z)]\ L_\pi(Z)\,M(XZ)\,,
\end{align}
and hence
\begin{align}
 L_\pi(Z)\,M(XZ)
   &= \sum_\lambda\ s_\lambda^{(\pi)}(X)\ s_\lambda(Z)\,.  
\end{align}

\mybenv{Example} 
\begin{align}
 \prod_{i\leq j} (1-z_iz_j)\ \prod_{i,j} (1-x_iz_j)^{-1} 
   &= \sum_\lambda\ s_\lambda^{(2)}(X)\ s_\lambda(Z)\,;
   \nonumber \\
   \displaybreak
 \prod_{i<j} (1-z_iz_j)\ \prod_{i,j} (1-x_iz_j)^{-1} 
   &= \sum_\lambda\ s_\lambda^{(1^2)}(X)\ s_\lambda(Z)\,;
   \nonumber \\
 \prod_{i\leq j\leq k} (1-z_iz_jz_k)\ \prod_{i,j} (1-x_iz_j)^{-1} 
   &= \sum_\lambda\ s_\lambda^{(3)}(X)\ s_\lambda(Z)\,;
   \nonumber \\   
 \prod_{i\neq j} (1-z^2_iz_j)\ \prod_{i<j<k} (1-z_iz_jz_k)^2\ 
    \prod_{i,j} (1-x_iz_j)^{-1} 
   &= \sum_\lambda\ s_\lambda^{(21)}(X)\ s_\lambda(Z)\,;
   \nonumber \\
 \prod_{i<j<k} (1-z_iz_jz_k)\ \prod_{i,j} (1-x_iz_j)^{-1}
   &= \sum_\lambda\ s_\lambda^{(1^3)}(X)\ s_\lambda(Z)\,.
\nonumber
\end{align}
\myeenv

In the first two cases, $s_\lambda^{(2)}(X)$ and $s_\lambda^{(1^2)}(X)$,
with the appropriate specification of $X$, are nothing other than
the irreducible orthogonal and symplectic group characters, variously
denoted by $o_\lambda(X)=[\lambda](X)$ and $sp_\lambda(X)=\la\lambda\ra(X)$,
respectively~\cite{littlewood:1950a,baker:1996a}. In the remaining cases,
$s_\lambda^{(3)}(X)$, $s_\lambda^{(21)}(X)$ and  $s_\lambda^{(1^3)}(X)$,
again with appropriate specifications of $X$, are formal, not necessarily
irreducible, characters of the subgroups $H_3(n)$, $H_{21}(n)$ and
$H_{1^3}(n)$ of $GL(n)$ that leave invariant cubic forms of symmetry specified
by the partitions $(3)$, $(21)$ and $(1^3)$,
respectively~\cite{fauser:jarvis:king:wybourne:2006a}.

\section{Vertex operators}\label{section:VertexOperators}
\setcounter{equation}{0}

\subsection{Vertex operators associated with formal characters}
\label{subsection:VertexOperatorsFormalCharacters}
Let $X$ be the underlying alphabet of all our Schur functions and Schur function
series unless otherwise indicated, with $X$ itself suppressed unless it is
necessary to exhibit it. In addition let $Z=(z_1,z_2,\ldots,z_l)$, and for any
partition $\lambda$ of length $\ell(\lambda)\leq l$ let $Z^\lambda = 
z_1^{\lambda_1}\,z_1^{\lambda_2}\,\cdots\,z_l^{\lambda_l}$ and let 
$[Z^\lambda\,](\cdots)$ be the coefficient of $Z^\lambda$ in $(\cdots)$. For any
non-zero $z$ let $\ov{z}=1/z$. Then the vertex operator $V(z)$ is defined by
\begin{align}
 V(z) &= M(z)\, L^\perp(\ov{z})\,.
\end{align}
If phrased in the language of symmetric functions, see for 
example~\cite[Ex.29 p.95]{macdonald:1995a} and ~\cite{cai:jing:2010a}, vertex operators 
are also sometimes called Bernstein vertex operators, or simply 
Bernstein operators, a name coined by Zelevinsky~\cite[p.69]{zelevinsky:1981a}.
The following formula is well-known~\cite{macdonald:1995a,baker:1996a}:
\medskip
\mybenv{Proposition}
Let $\lambda=(\lambda_1,\lambda_2,\ldots,\lambda_l)$ be a partition of length
$\ell(\lambda)\leq l$. Then
\begin{equation}
  s_\lambda = [Z^\lambda\,]\ V(z_1)\,V(z_2)\,\cdots\,V(z_l)\cdot 1\,.
\label{Eq-slambda-vert}
\end{equation}
\label{Prop-slambda-vert}
\myeenv

\noindent{\bf Proof:}
One way to see this is as follows:
\begin{align}
 [Z^\lambda\,]
   &\ V(z_1)\,V(z_2)\,\cdots\,V(z_l)\cdot 1\ 
   \nonumber \\
   &= [Z^\lambda\,]\ M(z_1)\, L^\perp(\ov{z_1})\ M(z_2)\, L^\perp(\ov{z_2})\ 
       \cdots\ M(z_l)\, L^\perp(\ov{z_l})\ \cdot 1\,.
\end{align}
However, since the outer coproduct of $L$ is just $\Delta L = L\otimes L$, we
have 
\begin{align}
 L^\perp(\ov{z})\,( M(w)\, G ) 
   &= (M(w)/L(\ov{z}))\ \ (G/L(\ov{z}))\,
\end{align}
for any $G$, $w$ and non-zero $z$, while
\begin{align}
 M(w)/L(\ov{z})
   &= M(w)/(s_0-\ov{z}s_1)= M(w)-\ov{z}\,w\,M(w).
\end{align}
Noting that $L^\perp(\ov{z}) \cdot 1 = 1$ for all non-zero $z$, this implies that
\begin{align}
 [Z^\lambda\,]
   &\ V(z_1)\,V(z_2)\,\cdots\,V(z_l)\cdot 1
   \nonumber \\ 
   &= [Z^\lambda\,]\ \prod_{1\leq i<j\leq l} (1-\ov{z_i}\,z_j) M(z_1)\, M(z_2)\, 
      \cdots\, M(z_l)
   \nonumber \\
   &= [Z^{\lambda+\delta}\,]\ \prod_{1\leq i<j\leq l} (z_i-z_j)\ M(Z) 
    = [s_\lambda(Z)]\, M(Z)\,,
\end{align}
where $\delta=(n-1,\ldots,1,0)$ and use has been made of the fact that
$$
s_\lambda= (Z^{\lambda+\delta}+\cdots)/\prod_{i<j}(z_i-z_j),
$$ 
while
$M(z_1)\, M(z_2)\, \cdots\, M(z_l)=M(Z)$. Restoring the $X$ dependence for
the moment,
\begin{align}
 M(XZ)
   &= \prod_{i=1}^n\prod_{j=1}^{l}\ (1- x_iz_j)^{-1} 
    = \sum_\lambda\ s_\lambda(X)\ s_\lambda(Z) \,,
\end{align}
so that $[s_\lambda(Z)]\,M(XZ)= s_\lambda(X)$. That is to say, without the
explicit $X$-dependence, we have $[s_\lambda(Z)]\, M(Z)=s_\lambda$, as required
to complete the proof of (\ref{Eq-slambda-vert}).
\qed

In order to generalise Proposition~\ref{Prop-slambda-vert} to the characters
$s_\lambda^{(\pi)}$ for arbitrary partitions $\pi$, it is helpful to first
establish
\mybenv{Lemma}
For all $w$, $z$ and all partitions $\pi$ of weight $|\pi|=p\geq 1$,
\begin{align}
 L_{\pi}^\perp(w)\, M(z)
   &= (1- w\,z^p\,\delta_{\pi,(p)})\, M(z)
      \,\prod_{k=1}^{p-1}\, L^\perp_{\pi/(k)}(wz^k)\,\, L_{\pi}^\perp(w)\,,
\label{Eq-Lpi-perp-M}
\end{align}
where the product over $k$ is absent if $p=1$, that is to say if $\pi=(1)$.
\label{Lem-Lpi-perp-M}
\myeenv

\noindent{\bf Proof:} For arbitrary $G$, 
\begin{align}
   L_{\pi}^\perp(w)\,(M(z)\, G) &= (M(z)\, G) / L_\pi(w)
   \nonumber \\
   &= M(z)/(L_\pi(w))_{(1)}\ \ \  G/(L_\pi(w))_{(2)} \,,
\end{align}
where, in Sweedler notation, 
\begin{align}
 \Delta L_\pi(w)
   &= (L_\pi(w))_{(1)} \otimes (L_\pi(w))_{(2)} \,.
\end{align}
More explicitly, in terms of Littlewood-Richardson coefficients,
\begin{align}
 \Delta L_{\pi}(w)  
   &= (\,L_\pi(w)\ \otimes L_\pi(w)\,) 
    \displaystyle{ \prod_{0\neq\xi,\eta\neq \pi} 
                  \prod_{c=1}^{c^\pi_{\xi\eta}}  
                  \sum_{\rho(\xi,\eta,c)}\ 
                   (-w)^{|\rho(\xi,\eta,c)|}\ s_{\rho(\xi,\eta,c)}[s_\xi]\ 
                        \otimes
                        s_{\rho(\xi,\eta,c)'}[s_\eta]}
    \nonumber \\
   &= (L_{\pi}(w))_{(1)} \otimes (L_{\pi}(w))_{(2)}\,.
\label{Eq-Delta-L_pi}
\end{align} 
However, since $M(z)=\sum_{m\geq0}\, s_{(m)}\ z^m$, we have 
\begin{align}
 M(z)/s_\lambda
   &= \left\{\begin{array}{cl}
               z^k\,M(z)&~~\hbox{if $\lambda=(k)$ with $k\geq0$};\\
               0&~~\hbox{otherwise}.
             \end{array} \right.
\label{Eq-M-skew-sf}
\end{align}
Moreover, for any partitions $\rho$ and $\xi$, the plethysm $s_\rho[s_\xi]$ is
such that
\begin{align}
 M(z)/(\,s_\rho[s_\xi]\,)
   &=  \left\{\begin{array}{cl}
                  z^{rk}\,M(z)&~~\hbox{if $\rho=(r)$ and $\xi=(k)$ with 
                     $r,k\geq 0$};     \\
                  0&~~\hbox{otherwise}.\\
              \end{array} \right.
\label{Eq-M-skew-spleth}
\end{align} 
It follows first from this that
\begin{align}
 M(z)/L_\pi(w)
   &= \sum_{r\geq0} (-w)^r  M(z)/(\,s_{1^r}[s_\pi]\,) 
    = (1-w\,z^p\, \delta_{\pi,(p)})\, M(z)\,.
\label{Eq-M-skew-Lpi}
\end{align}
Then in evaluating all other contributions of the form $M(z)/(L_\pi(w))_{(1)})$,
with $(L_\pi(w))_{(1)}$ identified as in (\ref{Eq-Delta-L_pi}), the domain of
$\xi$ may be restricted to one-part partitions $(k)$ with $0<k<p$, for which all
non-zero $c^\pi_{(k),\eta}$ are equal to $1$, so that $c=1$. In addition all
$\rho(\xi,\eta,c)$ may be restricted to partitions of the form $(r)$, with
$\rho(\xi,\eta,c)'=(1^r)$. If for fixed $\pi$ we let $m(k)$ be such that 
$s_{\pi/(k)}=\sum_{m=1}^{m(k)}\, s_{\eta(k,m)}$ for each $k=1,2,\ldots,p-1$, then
\begin{align}
 L_{\pi}^\perp(w) &\, (\, M(z)\, G\,)
   = \sum_{r(k,m)}\
       (\, M(z)/ (L_\pi(w)\, \prod_{k=1}^{p-1}\, 
        \prod_{m=1}^{m(k)}\ s_{(r(k,m))}[s_{(k)}] \,)\,)
   \nonumber \\
   \displaybreak
   &\hskip8em
       (\, G/ (L_\pi(w)\, \prod_{k=1}^{p-1}\, 
        \prod_{m=1}^{m(k)}\  (-w)^{r(k,m)}\  s_{(1^{r(k,m)})}[s_{\eta(k,m)}]\,)\,)\,.
   \nonumber 
\end{align}
Then from (\ref{Eq-M-skew-spleth}) it follows that
\begin{align}
 L_{\pi}^\perp(w)&\, (\, M(z)\, G\,)
  = \sum_{r(k,m)} 
       (\, \prod_{k=1}^{p-1}\, \prod_{m=1}^{m(k)}\  \,z^{k\,r(k,m)} \ 
        M(z)/ L_\pi(w) \,)
   \nonumber 
   \\
  &\hskip4em
       (\, G/ (L_\pi(w)\, \prod_{k=1}^{p-1}\, 
        \prod_{m=1}^{m(k)}\ (-w)^{r(k,m)}\  
        s_{(1^{r(k,m)})}[s_{\eta(k,m)}]\,)\,) 
   \nonumber\\
  &= \sum_{r(k,m)} 
       (\, M(z)/ L_\pi(w) \,)
   \nonumber \\
  &\hskip4em
       (\, G/ (L_\pi(w)\, \prod_{k=1}^{p-1}\, 
        \prod_{m=1}^{m(k)}\ (-w)^{r(k,m)}\,z^{k\,r(k,m)}\
	s_{(1^{r(k,m)})}[s_{\eta(k,m)}]\,)\,)\,.
   \nonumber  \\  
\end{align}
Hence
\begin{align}
 L_{\pi}^\perp(w) (\, M(z)\, G\,)
  &= (\, M(z)/ L_\pi(w) \,)\ \ \ 
     (\, G/ (L_\pi(w)\, \prod_{k=1}^{p-1}\, 
      \prod_{m=1}^{m(k)}\ L_{\eta(k,m)}(wz^k) \,) \,)
   \nonumber \\
  &= (\, M(z)/ L_\pi(w) \,)\ \ \
     (\, G/ (L_\pi(w)\, \prod_{k=1}^{p-1}\, L_{\pi/(k)}(wz^k) \,) \,)\,.
\end{align}
From (\ref{Eq-M-skew-Lpi}) it follows that for all $G$
\begin{align}
 L_{\pi}^\perp(w)\, (\, M(z)\, G\,) \
   &= (1-w\,z^p\,\delta_{\pi,(p)})\, M(z)\ (\, G/ (L_\pi(w)\, \prod_{k=1}^{p-1}\, L_{\pi/(k)}(wz^k) \,) \,)  \,,
\end{align}
which implies the validity of (\ref{Eq-Lpi-perp-M}).
\qed

This Lemma leads immediately to the following generalization of Proposition~\ref{Prop-slambda-vert}
\mybenv{Proposition}\label{Prop-slambda-pi-vert}
Let $\lambda=(\lambda_1,\lambda_2,\ldots,\lambda_l)$ be a partition
of length $\ell(\lambda)\leq l$, and let $\pi$ be a partition of weight
$|\pi|=p\geq1$. Then 
\begin{align}
 s_\lambda^{(\pi)}
   &= [Z^\lambda\,]\ V^\pi(z_1)\,V^\pi(z_2)\,\cdots\,V^\pi(z_l)\cdot 1\,,
\label{Eq-slambda-pi-vert}
\end{align}
where 
\begin{align}
 V^\pi(z)
   &= (1-z^p\,\delta_{\pi,(p)})\, M(z)\ L^\perp(z^{-1})\  \prod_{k=1}^{p-1}\, L^\perp_{\pi/(k)}(z^k) \,.
\end{align}
\myeenv

\noindent{\bf Proof}: Proceeding as in the proof of 
Proposition~\ref{Prop-slambda-vert}, one can make use of the fact that 
\begin{align}
 L^\perp(\ov{z})
   &\,( (1-w^p\,\delta_{\pi,(p)})\,M(w)\, G )  
    =  L^\perp(\ov{z})\ (\,(M(w)/L_\pi)\ G )
   \nonumber \\
\displaybreak   
   &= M(w)/(L_\pi\,L(\ov{z})\ \ (G/L(\ov{z}))
    = (1-w\,\ov{z})\,(M(w)/L_\pi)\ (G/L(\ov{z}))
   \nonumber \\
   &= (1-w\,\ov{z})\, (1-w^p\,\delta_{\pi,(p)})\,M(w)\ (G/L(\ov{z}))\,, 
\end{align}
for any $G$, $w$ and non-zero $z$, to show that
\begin{align}
 [Z^\lambda\,] 
   &\ V^\pi(z_1)\,V^\pi(z_2)\,\cdots\,V^\pi(z_l)\cdot 1
   \nonumber \\ 
   &= [Z^\lambda\,]\ \prod_{1\leq i<j\leq l} (1-\ov{z_i}\,z_j)\ 
         \prod_{k=1}^l\ (1-z_k^p\,\delta_{\pi,(p)})
   \nonumber \\
   &\hskip2em M(z_1)\  \prod_{k=1}^{p-1}\, L^\perp_{\pi/(k)}(z_1^k) \
       \cdots  M(z_l)\ \prod_{k=1}^{p-1}\, L^\perp_{\pi/(k)}(z_l^k) 
       \cdot 1
   \nonumber \\
   &= [Z^{\lambda+\delta}\,]\ \prod_{1\leq i<j\leq l} (z_i-z_j)\ 
       \prod_{k=1}^l\ (1-z_k^p\,\delta_{\pi,(p)})
   \nonumber \\
   &\hskip2em  M(z_1)\ \prod_{k=1}^{p-1}\, L^\perp_{\pi/(k)}(z_1^k)\ 
       \cdots  M(z_l)\ \prod_{k=1}^{p-1}\, L^\perp_{\pi/(k)}(z_l^k)\cdot 1
   \nonumber \\
   &= [s_\lambda\,]\  \prod_{k=1}^l\ (1-z_k^p\,\delta_{\pi,(p)})
   \nonumber \\
   &\hskip2em  M(z_1)\ \prod_{k=1}^{p-1}\, L^\perp_{\pi/(k)}(z_1^k)\
       \cdots  M(z_l)\ \prod_{k=1}^{p-1}\, L^\perp_{\pi/(k)}(z_l^k)
       \cdot 1
   \nonumber \\
   &= [s_\lambda\,]\ \ L^\perp_\pi\, M(z_1)\,\cdots\, M(z_l) \cdot 1\ 
    = [s_\lambda]\ L^\perp_\pi\,(M(Z)) \ =\  s^\pi_\lambda \,,
\label{Eq-Prop-proof}
\end{align}
as required, where the first step in the last line involves the use of (\ref{Eq-Lpi-perp-M}) extended iteratively,
and the final step is a consequence of (\ref{Eq-slambda-pi}).
\qed

\subsection{Vertex operators in exponential form} 
Given  
\begin{align}
 M(z;X) 
   &=\prod_{i\geq1} (1-z\,x_i)^{-1}\,,
\end{align} 
it follows that 
\begin{align}
 \ln M(z;X) 
  &= {-\sum_{i\geq1} \ \ln (1-z\,x_i) 
   = \sum_{i\geq1}\ (z\,x_i + \frac{(z\,x_i)^2}{2} +\frac{(z\,x_i)^3}{3} 
      + \cdots ) }
  \nonumber \\
  &= { z\,p_1(X) + \frac{z^2}{2}\, p_2(X) + \frac{z^3}{3}\, p_3(X) +\cdots } 
   = \sum_{k\geq1} \ \frac{z^k}{k}\, p_k(X)\,.
\end{align}
Hence
\begin{align}
 M(z;X)
   &= \exp \left( \sum_{k\geq1} \ \frac{z^k}{k}\, p_k(X) \right)\,.
\end{align}

It follows that for any partition $\pi$
\begin{align} 
 \ln M_\pi(z;X)
   &= \sum_{k\geq1} \ \frac{z^k}{k}\, p_k(Y) 
    = \sum_{k\geq1} \ \frac{z^k}{k}\, p_k[s_\pi(X)]
    =   \sum_{k\geq1} \ \frac{z^k}{k}\, s_\pi[p_k(X)]
\displaybreak
   \nonumber \\
   &= \sum_{k\geq1} \ \frac{z^k}{k}\, \left(\sum_\rho \frac{1}{z_\rho} 
        \chi^\pi_\rho\ p_\rho[p_k(X)]\right)
    = \sum_{k\geq1} \ \frac{z^k}{k}\, \left(\sum_\rho \frac{1}{z_\rho}
        \chi^\pi_\rho\ p_{k\rho}(X)\right)\,,
\end{align}
where we have used the identities $p_k[s_\pi(X)] = s_\pi[p_k(X)]$ and $p_r[p_k(X)]=p_{rk}(X)$
that apply for all $\pi$, $k$, $r$ and $X$, and the notation $k\rho=(k\rho_1,k\rho_2,\ldots)$ if  
$\rho=(\rho_1,\rho_2,\ldots)$. 

\mybenv{Example}\label{Ex-lnM}
Suppressing the $X$ dependence
\begin{align}
 \ln M_{(1)}(z) 
   &= \sum_{k\geq1} \ \left(\, p_{k}\,\right) z^k/k \,;
   \nonumber \\
 \ln M_{(2)}(z)
   &= \sum_{k\geq1} \ \left(\, p_{(k,k)}+p_{(2k)}\,\right) z^k/2k \,;
   \nonumber \\
 \ln M_{(1,1)}(z)
   &= \sum_{k\geq1} \ \left(\, p_{(k,k)}-p_{(2k)}\,\right) z^k/2k \,;
   \nonumber \\
 \ln M_{(3)}(z)
   &= \sum_{k\geq1} \ \left(\, p_{(k,k,k)}+3p_{(2k,k)}+2p_{(3k)}\,\right) 
        z^k/6k \,;
   \nonumber \\
 \ln M_{(2,1)}(z)
   &= \sum_{k\geq1} \ \left(\,p_{(k,k,k)}-p_{(3k)}\,\right) z^k/3k \,;
   \nonumber \\
 \ln M_{(1,1,1)}(z)
   &= \sum_{k\geq1} \ \left(\, p_{(k,k,k)}-3p_{(2k,k)}+2p_{(3k)}\,\right) 
        z^k/6k \,.
\end{align}
\myeenv
Since  
\begin{align}
      L(z;X) &=\prod_{i\geq1} (1-z\,x_i)= 1/M(z;X) \,,
\end{align} 
it follows that
\begin{align}
 \ln L(z;X) 
   &= - \ln M(x;Z)
    = - \sum_{k\geq1} \ \frac{z^k}{k}\, p_k(X)\,,
\end{align}
and more generally
\begin{align}
 \ln L_\pi(z;X)
   &= - \ln M_\pi(x;Z)
    = - \sum_{k\geq1} \ \frac{z^k}{k}\, \left(\sum_\rho \frac{1}{z_\rho} 
              \chi^\pi_\rho\ p_{k\rho}(X)\right)\,.
\end{align}
Then, if we recall that for all positive integers
$k$~\cite[p.76]{macdonald:1995a},
\begin{align}
 p^\perp_k(X)
   &= k \frac{\partial}{\partial p_k(X)}\,,
\end{align}
we are in a position to see that
\begin{align}
 L^\perp(z;X)
   &= L_{(1)}^\perp(z)
    = \exp \left( -\sum_{k\geq1} \ z^k\, \frac{\partial}{\partial p_k(X)}
           \right)\,,
\label{Eq-Lperp1}     
\end{align}
\vfill
\newpage
while, exploiting the data of Example~\ref{Ex-lnM}, we have:
\begin{align}
 L_{(2)}^\perp(z;X)
   &= \exp \left(- \sum_{k\geq1}\ z^k\, \left( \frac{k}{2}\, 
         \frac{\partial^2}{\partial p_k(X)^2}
      +  \frac{\partial}{\partial p_{2k}(X)} \right)\right)\,;
   \nonumber \\
 L_{(1^2)}^\perp(z;X)
   &= \exp \left(- \sum_{k\geq1}\ z^k\, \left( \frac{k}{2}\, 
         \frac{\partial^2}{\partial p_k(X)^2}
       - \frac{\partial}{\partial p_{2k}(X)} \right)\right)\,.
\label{Eq-Lperp2}     
\end{align}

Proposition~\ref{Prop-slambda-pi-vert} implies:
\begin{align}
 V^{(1)}(z)
   &= (1-z)\, M(z)\, L^\perp(z^{-1})\,;
   \nonumber \\
 V^{(2)}(z)
   &= (1-z^2)\, M(z)\, L^\perp(z^{-1})\, L^\perp_{(1)}(z)\,;
   \nonumber \\
 V^{(1^2)}(z)
   &= M(z)\, L^\perp(z^{-1})\, L^\perp_{(1)}(z)\,;
   \nonumber \\
 V^{(3)}(z)
   &= (1-z^3)\, M(z)\, L^\perp(z^{-1})\, L^\perp_{(2)}(z)\,L^\perp_{(1)}(z^2)\,;
   \nonumber \\
 V^{(21)}(z)
   &= M(z)\, L^\perp(z^{-1})\, L^\perp_{(2)}(z)\,L^\perp_{(1^2)}(z)\,
             L^\perp_{(1)}(z^2)\,;
   \nonumber \\
 V^{(1^3)}(z)
   &=  M(z)\, L^\perp(z^{-1})\, L^\perp_{(1^2)}(z)\,.
\label{Eq-Vz-examples}   
\end{align}
The results (\ref{Eq-Lperp1}) and (\ref{Eq-Lperp2}) are then sufficient 
for us to express the vertex operators of (\ref{Eq-Vz-examples})
in exponential form as follows:
\begin{align}
 V^{(1)}(z)
   &= \displaystyle(1-z)\ \exp\left( 
         \sum_{k\geq1} \frac{z^k}{k}\, p_{k}\,\right) 
         \exp\left(-\sum_{k\geq1}\, z^{-k}\, \frac{\partial}{\partial p_k} 
	 \right) ;
   \nonumber \\
 V^{(2)}(z)
   &= \displaystyle(1-z^2)\ \exp\left( \sum_{k\geq1} \frac{z^k}{k}\, 
         p_{k}\,\right) 
         \exp\left(-\sum_{k\geq1} 
         (z^{-k}+z^k) \frac{\partial}{\partial p_k} \right) ;
   \nonumber \\
 V^{(1^2)}(z)
   &=\displaystyle \exp\left( \sum_{k\geq1} \frac{z^k}{k}\, p_{k}\,\right) 
         \exp\left(-\sum_{k\geq1} (z^{-k}+z^k) \frac{\partial}{\partial p_k} 
	 \right) ;
   \nonumber \\
 V^{(3)}(z)
   &= \displaystyle(1-z^3)\ \exp\left( \sum_{k\geq1} \frac{z^k}{k}\, 
         p_{k}\,\right)
   \nonumber \\
   &\phantom{+++++}\displaystyle \exp\left( -\sum_{k\geq1} \ 
         \left((z^{-k}+z^{2k})\frac{\partial}{\partial p_k} 
             + \frac{k\,z^k}{2}\, \frac{\partial^2}{\partial p_k^2}
             + z^k\, \frac{\partial}{\partial p_{2k}} \right)  \right);
   \nonumber \\
 V^{(21)}(z)
   &= \displaystyle\ \exp\left( \sum_{k\geq1} \frac{z^k}{k}\, p_{k}\,\right)\
         \exp\left( -\sum_{k\geq1} \ \left((z^{-k}+z^{2k})
         \frac{\partial}{\partial p_k} 
              + k\,z^k\, \frac{\partial^2}{\partial p_k^2}
              \right)  \right);
   \nonumber \\
 V^{(1^3)}(z)
   &= \displaystyle\ \exp\left( \sum_{k\geq1} \frac{z^k}{k}\, p_{k}\,\right)
                     \exp\left( -\sum_{k\geq1} \ \left(z^{-k}
                      \frac{\partial}{\partial p_k} 
                    + \frac{k\,z^k}{2}\, \frac{\partial^2}{\partial p_k^2}
                    - z^k\, \frac{\partial}{\partial p_{2k}} \right)  \right),
\label{Eq-VertexOperator-table}
\end{align}
where once again the explicit dependence on $X$ has been omitted.

The first result expresses the fact that $V^{(1)}(z)=(1-z)\,V(z)$. At first
sight this appears rather surprising, but it should be noted that it yields 
\begin{align}
 s^{(1)}_\lambda 
   &= [Z^\lambda]\ V^{(1)}(z_1)\,V^{(1)}(z_2)\,\cdots\,V^{(1)}(z_l)\cdot 1
   \nonumber \\
   &= [Z^\lambda\,]\ \prod_{i=1}^l (1-z_i)\ \, V(z_1)\,V(z_2)\,\cdots\,V(z_l)
     \cdot 1
   \nonumber \\ 
   &= [s_\lambda(Z)]\ \prod_{i=1}^l (1-z_i)\ M(Z)\,.
\end{align}
Implicit in this is the dependence on an arbitrary alphabet
$X=(x_1,x_2,\ldots\,)$. Making this explicit gives
\begin{align}
 s^{(1)}_\lambda(X)
   &= [s_\lambda(Z)]\ \prod_{i=1}^l (1-z_i)\ M(XZ)
    = [s_\lambda(Z)]\ L(Z)\ M(XZ)
   \nonumber \\
   &= [s_\lambda(Z)]\ L^\perp(X)\,(M(XZ))
    = [s_\lambda(Z)]\ L^\perp(X)\left(\sum_\mu\,s_\mu(X)\,s_\mu(Z)\right)
   \nonumber \\
   &= L^\perp(X)\,(s_\lambda(X))
    = L_{(1)}^\perp\,(s_\lambda(X))\,, 
\end{align} 
as required by the definition (\ref{Eq-def-slambda-pi}) of such a character.

The next two results in (\ref{Eq-VertexOperator-table}) have been derived by
Baker~\cite{baker:1996a} using different techniques involving rather more
traditional operator reordering methods. In what follows next, this operator 
ordering approach is outlined and is used, by way of example, to
recover the formula for $V^{(21)}$ as given in (\ref{Eq-VertexOperator-table}).
However, it is clear that our Proposition~\ref{Prop-slambda-pi-vert} allows
further vertex operators $V^{\pi}$ specified by partitions $\pi$ of weight
higher than $3$ to be written down rather easily.  

\subsection{Vertex operators via normal ordering}
\label{subsection:OperatorOrdering}
The expressions $L_\pi^\perp(w;X)(M(XZ))$ can be evaluated rather easily
using~(\ref{Eq-FM-FperpM}), or more explicitly for any given $\pi$ as in 
Appendix~\ref{appendix:LperpM}. 
However, from the point of view of operator ordering the more
general expressions $L_\pi^\perp(w;X)M(XZ)$ are not in so-called normal-ordered
form since they involve exponentials of various partial derivatives with
respect to power sum symmetric functions standing to the left of other
exponentials of power sum symmetric functions. 

Algebraically, if we introduce operators $K$ and $P$ such that 
$e^K = L_\pi^\perp(w;X)$ and $e^P = M(XZ)$, the normal ordering 
problem can be tackled using the following formula: \\
\begin{align}
 e^K e^P 
   &= \,  e^P\big( e^{-P} e^K e^P\big)
    = e^P \ e^{\,K\ +\ [K,P]\ +\ \frac 12[[K,P],P]\ +\ \frac 16[[[K,P],P],P]\ 
        +\ \cdots\ }\,.
\label{Eq-eKeP}
\end{align}
Even though we suspect the formula used in making the second step 
may be well known as an adjoint action result
in the theory of Lie groups and their algebras, we have been unable to 
locate a statement or proof of this result. We therefore attach a strictly
combinatorial proof in Appendix~\ref{appendix:bchd}. 

In our case, $K$ symbolises the partial derivative operator of degree $p$ that
is defined by $L_\pi^\perp(w;X)=\exp K$ with $p=\vert \pi\vert$, and $P$ is  
the usual infinite sum of power sum symmetric functions appearing as the
exponent in the formula $M(XZ)=\exp P$. One must retain terms up to those
involving $1/p!$ in the expansion of the exponent that appears
in~(\ref{Eq-eKeP}) in order to extract all scalar and differential
contributions arising from the reordering. Note that none of the surviving
terms in the exponent of the final term will contain symmetric functions in the alphabet $X$,
but only scalars and partial derivatives that all mutually commute. This
enables this term to be written as a product of exponentials, each with a single
multi-commutator argument. 

As an illustration of this method, we deal with the case $\pi=(21)$ for which
$p=3$. For ease of writing, we suppress the alphabet $X$ and abbreviate
$\partial/\partial p_k(X)$ as $\partial_k$ for all positive integers $k$. 
In this case
\begin{align}
 M_{(21)}(w) 
   &= \, \exp \left( \sum_{k\geq1} \frac{w^k}{3k}(- p_{3k}+ p_k^3) \right); 
   &&& 
 L_{(21)}(w)
   &= \exp \left( \sum_{k\geq1} \frac{w^k}{3k} (p_{3k} - p_k^3) \right); 
   \nonumber \\
 L^\perp_{(21)}(w) 
   &= \, \exp \left( \sum_{k\geq1} w^k\,( 
     \partial_{3k}- \frac{1}{3} k^2 \partial_k^3) \right);
   &&&
 M(Z)
   &= \exp \left(\sum_{m\geq1} \frac {p_m\cdot p_m(Z)}{m} \right)\,.
\end{align}
This gives $K =  \sum_{k\geq1}w^k \left( \partial_{3k}- \frac{1}{3}k^2 \partial_k^3\right)$ and
$P = \sum_{m\geq1} {p_m\cdot p_m(Z)}/{m}$. We calculate directly
\begin{align}
 {[}\partial_{3k},P{]} 
  &= \frac {1}{3k}\,p_{3k}(Z)\,;
  \nonumber \\
 {[}k^2 \partial_k^3,P{]}
  &= 3k\, p_k(Z)\, \partial_k^2 ;
  \nonumber \\
 {[}{[}k^2 \partial_k^3,P{]},P{]}
  &=  6\, p_k(Z)^2\, \partial_k ;
 \nonumber \\
 {[}{[}{[}k^2 \partial_k^3,P{]},P{]},P{]}
  &= \frac{6}{k}\, p_k(Z)^3\,.
\end{align}
Hence
\begin{align}
  [K,P]= \sum_{k\geq1} w^k\left(\frac{1}{3k}p_{3k}(Z) 
                     - k\,p_k(Z)\,\partial_k^2\right)\,, 
\end{align}
which clearly commutes with $K$ as claimed. Moreover,
\begin{align}
K &+
 {[}K,P{]} + \frac 12 {[}{[}K,P{]},P{]} + \frac 16 {[}{[}{[}K,P{]},P{]},P{]} 
  \nonumber \\
  &= \sum_{k\geq1}w^k\left(\frac{1}{3k}\big( p_{3k}(Z)-p_k(Z)^3\big) 
     + \partial_{3k} - p_k^2(Z)\,\partial_k - k\,p_k(Z)\,\partial_k^2 
     - \frac{1}{3} k^2 \partial_k^3 \right). 
\end{align}
It follows that
\begin{align}
L_{(21)}^\perp(w)\,M(Z)
  &= M(Z) \, L_{(21)}(w;Z) \, \exp \left(- \sum_{k\geq1} w^k 
        \big(p_k^2(Z)\,\partial_k + k\,p_k(Z)\,\partial_k^2 \big) \right) 
        \, L_{(21)}^\perp(w)\,.
\label{Eq-LperpM-21}
\end{align}
\vfill
\newpage

There are a number of special cases of this normal ordered formula
that are of interest. First, acting on $1$, or any other scalar, this gives
\begin{align}
 L_{(21)}^\perp(w)\,M(Z)\cdot 1 
   &= \,  M(Z) \, L_{(21)}(w;Z) \cdot 1
\end{align}
in agreement with the identity (\ref{Eq-FM-FperpM}).

Second, restricting $Z$ to the one letter alphabet $z$ gives
\begin{align}
 p_k(z)
   &=z^k, \quad p_{2k}(z)=z^{2k}
   &&\hbox{and}&&&
 p_{3k}(z)-p_k(z)^3
   &=z^{3k}-(z^k)^3=0\,,
\end{align}
for all $k\geq1$, so that $L_{(21)}(z)=1$, and (\ref{Eq-LperpM-21}) reduces to
\begin{align}
  L_{(21)}^\perp(w)\,M(z)
    &= M(z) \, \exp \left( - \sum_{k\geq1} w^k
        \big(z^{2k}\,\partial_k + k z^k \,\partial_k^2\big) \right)
        \, L_{(21)}^\perp(w) \,.
\label{Eq-LperpM-Mpartial}
\end{align}

This is nothing other than an illustrative example of
Lemma~\ref{Lem-Lpi-perp-M} since the identity
\begin{align}
 z^{2k}\,\partial_k + k z^k \,\partial_k^2
  &= \big(\frac{1}{2} k z^k\, \partial_k^2 + z^k \partial_{2k}\big) 
   + \big(\frac{1}{2} k z^k\, \partial_k^2 - z^k \partial_{2k}\big)
   + z^{2k}\, \partial_k 
\end{align}
enables (\ref{Eq-LperpM-Mpartial}) to be rewritten in the form
\begin{align}
 L_{(21)}^\perp(w)\,M(z)
   &= M(z)\,L_{(2)}^\perp(wz)\, L_{(1^2)}^\perp(wz)\, L_{(1)}^\perp(wz^2)
      \, L_{(21)}^\perp(w)
   \nonumber \\
   &= M(z)\,\sum_{k=1}^2\ L_{(21/k)}^\perp(wz^k)\ L_{(21)}^\perp(w)\,,
\label{Eq-LperpM-21skewM}
\end{align}
where use has been made of (\ref{Eq-Lperp1}) and (\ref{Eq-Lperp2}).

In addition, it follows from~(\ref{Eq-LperpM-Mpartial}) that if we now set
$w=1$ and $Z=(z_1,z_2,\ldots,z_l)$ once again and reverse the sequence of
steps used in~(\ref{Eq-Prop-proof}) we obtain
\begin{align}
 s^{(21)}_\lambda
   &= [s_\lambda]\ L_{(21)}^\perp\,M(Z)\cdot 1 
    = [Z^\lambda]\ V^{(21)}(z_1)\ V^{(21)}(z_2)\ \cdots\ 
        V^{(21)}(z_l)\cdot 1 \,,  
\end{align}
with
\begin{align}
 V^{(21)}(z)
   &= M(z) \, \exp \left( - \sum_{k\geq1} 
       \big(z^{2k}\,\partial_k + k z^k \,\partial_k^2\big) \right)
   \nonumber \\
   &= \exp \left( \sum_{k\geq1} \frac{z^k}{k}\, p_k \right) \, \exp 
       \left( - \sum_{k\geq1} 
       \big( (-z^{-k}+z^{2k})\,\partial_k + k z^k \,\partial_k^2\big) \right),
\end{align}
precisely as in (\ref{Eq-VertexOperator-table}).

The other results of (\ref{Eq-VertexOperator-table}) may be obtained in the
same way.
\medskip

\section{Replicated vertex operators}\label{section:ReplicatedVertexOperators}
\setcounter{equation}{0}

Since their introduction in string theory, vertex operators have played a
fruitful role in mathematical constructions of group representations as well
as combinatorial objects. We cite for example applications to affine Lie 
algebras~\cite{lepowsky:wilson:1978a,frenkel:kac:1980a}, quantum affine
algebras~\cite{frenkel:jing:1988a} and sporadic discrete
groups~\cite{frenkel:lepowsky:meurman:1988a}, see
also~\cite[Chapter~14]{kac:1990a}. Variations on the theme of symmetric
functions~\cite{macdonald:1995a} are applications 
to $Q$-functions~\cite{jing:1991a,salam:wybourne:1992a}, Hall-Littlewood
functions~\cite{jing:1991b}, Macdonald
functions~\cite{macdonald:1988a,macdonald:1995a,jing:jozefiak:1992a,etingof:kirillov:1995a},
Jack functions~\cite{cai:jing:2010a} and Kerov symmetric
functions~\cite[Chapter~6]{baker:1994a}.

As a modest approach to generalising the vertex operators of
Section~\ref{section:VertexOperators}, the observations made in
Section~\ref{section:ParametrizedPlethysms} allow us to write down immediately
expressions for replicated or parameterized vertex operators. In the simplest
case, this is exemplified by
\begin{align}
 V_\alpha(z)
   &= M(\alpha z)\, L^\perp(\alpha z^{-1})
   \nonumber \\
   &= \exp\left(  \alpha \sum_{k\geq1}\frac{z^k}{k}\, p_{k}\,\right) 
      \exp  \left(-\alpha\sum_{k\geq1}\, z^{-k}\, \frac{\partial}{\partial p_k} 
      \right) \,.
\label{Eq-VOalpha}             
\end{align}
for any $\alpha$, integer, rational, real or complex. Here, making the usual
dependence on $X$ quite explicit,
\begin{align}
 M(\alpha z;X)
   &= M(z;X)^\alpha = \prod_{i\geq1} (1-z\,x_i)^{-\alpha}
   \nonumber \\
   &= \sum_{\sigma}\, s_\sigma(\alpha z)\, s_\sigma(X)
   \nonumber \\
   &= \sum_{\sigma}\, z^{|\sigma|} \dim_\sigma(\alpha)\ s_\sigma(X)\,,              
\end{align}
while
\begin{align}
 L(\alpha z^{-1};X)
   &= L(z^{-1};X)^\alpha = \prod_{i\geq1} (1-z^{-1}\,x_i)^\alpha
   \nonumber \\
   &= \sum_{\tau}\, (-1)^{|\tau|}\, s_{\tau}(\alpha z^{-1})\, s_{\tau'}(X)
   \nonumber \\
   &= \sum_{\tau}\, (-z)^{-|\tau|} \dim_{\tau}(\alpha)\ s_{\tau'}(X)\,,
\end{align}
as given first in~\cite{jarvis:yung:1993a}.

More generally, we can define in a similar way:
\begin{align}
 V^\pi_\alpha(z)
   &= (1-z^p\,\delta_{\pi,(p)})^\alpha\, M(\alpha z)\, L^\perp(\alpha z^{-1})
      \prod_{k=1}^{p-1} L^\perp_{\pi/(k)}(\alpha z^k)\,,
\end{align} 
where each term on the right has an expansion of the type shown above involving
sums over partitions $\sigma,\tau,\ldots$. Still more variations may be 
constructed in which the $\alpha$'s on the right are not all identical. It should be stressed that the
normal ordering relations for products of such vertex operators are very much
more complicated than those encountered in
Section~\ref{section:VertexOperators}.

\medskip
\section{Conclusion}\label{section:Conclusion}
\setcounter{equation}{0} 

This work allows us to conclude that Littlewood's attempt to
develop character theory algebraically, instead of using a group manifold
integration approach, leads much further than expected. In particular,
it allows us to obtain generating functions for formal characters of
a range of subgroups $H_\pi$ of the general linear group, well beyond the classical
orthogonal and symplectic subgroups. Furthermore, we see that the algebraic 
approach does not suffer from the infinities encountered using analytic
methods. Instead, the algebraic infinities are just those associated with 
readily manipulated infinite series of Schur functions. A further advantage 
of this approach is that the results of these manipulations take very compact 
forms if carried out in Hopf algebraic terms, as illustrated in the proof
of Proposition~\ref{Prop-slambda-pi-vert}, by way of Lemma~\ref{Lem-Lpi-perp-M},
that was based on a knowledge of the coproduct of an infinite Schur function
series. While the explicit exponential form of vertex operators cannot easily
yield such general results, it can be used to express the algebraic results
in a physically more desirable form, as exemplified
in~(\ref{Eq-VertexOperator-table}). Moreover, thanks to the plethystic
approach to replication and parameterization, these exponential forms can be
readily generalized, while still remaining susceptible to actual (machine)
calculations.

Moreover we have shown in Table~\ref{table1} that the use of inner
coproducts and the dimension map can dramatically speed up the computation of 
plethysms. This approach has two major benefits. It allows us i) to compute
plethysms with large multiplicities, for example $s_\mu[n\,s_\nu]$ for large
integers $n$, and ii) to extend plethysms to those involving an argument 
which need not be integral, but can be in a ring extension, and evaluate them. 
This opens the way for dealing with $q$-deformations, as introduced by Jarvis and 
Yung~\cite{jarvis:yung:1993a}, Baker~\cite{baker:1994a} and Brenti~\cite{brenti:2000a}, 
who considered plethysms of the form $s_\mu[q\,s_{(1)}]$, which correspond to scalings. 
In Appendix~\ref{appendix:code}
we provide pseudo code to implement an algorithm for their evaluation.

An extension of algebraic and operator methods in combinatorial settings,
which we have not pursued in the present work, invokes the fermion-boson
correspondence (see for
example~\cite{sato:miwa:jimbo:1977,date:kashiwara:miwa:1981a,jimbo:miwa:1983a}).
In the present case, our explicit vertex operator constructions for the
formal $H_\pi$ characters $s_\lambda^{(\pi)}$ can be expected to have their
equivalents in terms of free fermions, and hence via Wick's theorem, to be
amenable to determinantal evaluations~\cite{jarvis:yung:1994a}. The resulting
determinantal expressions can be expected to be helpful in an attack on the
notorious problem of determining the modification rules for $H_\pi$
characters involving a finite alphabet~\cite{fauser:jarvis:king:wybourne:2006a}.
We leave further developments along these lines to future work.

\medskip
\section*{Acknowledgement}

This work is one of a number of outcomes stimulated by our collaboration with
the late Professor Brian G Wybourne on the
paper~\cite{fauser:jarvis:king:wybourne:2006a}.  BF gratefully acknowledges the
Alexander von Humboldt Stiftung for {\it sur place}  travel grants to Hobart to
visit the School of Mathematics and Physics, University of Tasmania, and the
University of Tasmania for an honorary Research Associate appointment. Likewise
PDJ acknowledges longstanding support from the Alexander von Humboldt
Foundation, in particular for visits to the Max Planck Institute for Mathematics
in the Sciences, Leipzig. RCK acknowledges support from the Leverhulme
Foundation by way of an Emeritus Fellowship and travel grants enabling him to
visit the University of Tasmania. We thank the DAAD and the Emmy-Noether Zentrum
f\"ur Algebra at the University of Erlangen for the financial support of a visit
by RCK and PDJ to Erlangen in Spring 2009. This work could not have been
completed without a 'Research in Pairs' grant from the Mathematisches
Forschungsinstitut Oberwolfach, in Spring 2010, and all three authors wish to
record their appreciation of this award. Finally, it is a pleasure to
acknowledge the warm hospitality extended to the authors at all the above named
institutions in Hobart, Leipzig, Erlangen and Oberwolfach.

%
%
\begin{appendix}
\renewcommand{\theequation}{\Alph{section}-\arabic{equation}}

%
%
\section{An explicit expression for $L_\pi^\perp(w)\,(M(Z))$}
\label{appendix:LperpM}
\setcounter{equation}{0}

\mybenv{Proposition}
For any partition $\pi$, any $w$ and any alphabet $Z=(z_1,z_2,\ldots,z_l)$
with $l$ a positive integer,
\begin{align}
  L_\pi^\perp(w)\,(M(Z))
    &= \prod_{T\in{\cal T}^\pi[l]}\ (1-w\,Z^{\wgt(T)}) \ M(Z)\,,
\label{Eq-Lpi-perpw-M}
\end{align}
where the product is taken over all tableaux, $T$, in the set,
${\cal T}^\pi[l]$, of all semistandard or column-strict tableaux~\cite{macdonald:1995a} 
with entries taken from the
set $\{1,2,\ldots,l\}$. For each tableau $T$ its weight is defined to be
$\wgt(T)=(\#1,\#2,\ldots,\#l))$, with $\#k$ the number of entries $k$ in $T$
for $k=1,2,\ldots,l$.
\myeenv

\noindent{\bf Proof:}   
The proof is by induction with respect to $p$, the weight of $\pi$.

For $p=1$ we have $\pi=(1)$ and $L_\pi=L_{(1)}=L$ so that
\begin{align}
 L_{(1)}^\perp(w)\,(M(z))
  &= L^\perp(w)\,(\,M(z_1)\,M(z_2)\,\cdots\,M(z_l)\,)
  \nonumber \\
  &= M(z_1)/L(w)\ M(z_1)/L(w)\ \cdots\ M(z_l)/L(w)
  \nonumber \\
  &= (1-w\,z_1)(1-w\,z_2)\,\cdots\,(1-w\,z_l)\ M(z_1)\,M(z_2)\,\cdots\,M(z_l)
  \nonumber \\
  &= \displaystyle \prod_{1\leq i\leq l}  (1-w\,z_i) \ M(Z)\  
   = \prod_{T\in{\cal T}^{(1)}[l]}\ (1-w\,Z^{\wgt(T)}) \ M(Z)\,,
\end{align}
since each semistandard tableau of shape $\pi=(1)$ consists of a single box
whose entry is to be taken from $\{1,2,\ldots,l\}$. This proves the required
result in the case $p=1$.

Now we assume the result to be true for all partitions $\eta$ with weight
$|\eta|<p$. Then by means of the coproduct argument used in the proof of
Lemma~\ref{Lem-Lpi-perp-M} we have
\begin{align}
 L_{\pi}^\perp(w)\,(M(Z))
   &= L_\pi^\perp(w)\,(\,M(z_l)\cdots \,M(z_2)\,M(z_1)\,)
   \nonumber \\
   &= (\, M(z_l)/ L_\pi(w) \,)\ (\, M(Z')/ (L_\pi(w)\,
      \prod_{k=1}^{p-1}\, \prod_{m=1}^{m(k)}\ L_{\eta(k,m)}(w\,z_l^k) \,) \,) \,,
\label{Eq-Lpi-perp}      
\end{align}
where $Z'=(z_1,z_2,\ldots,z_{l-1})$. 

Consider the first factor $M(z_l)/L_\pi(w)$. The use of~(\ref{Eq-M-skew-Lpi}) 
implies that $M(z_l)/L_\pi(w)=M(z_l)$ unless $\pi=(p)$ in which case one
obtains
\begin{align}
  M(z_l)/L_{(p)}(w) 
    &= (1-w\,z^p)\, M(z_l) = (1- w\,z_l^{\wgt(T)})\ M(z_l)\,,
\end{align}
where $T$ is the single semistandard tableau of one-rowed shape $(p)$ whose 
entries are all $l$.

Turning to the second factor involving $M(Z')$, by the induction hypothesis
\begin{align}
  M(Z')/L_{\eta(k,m)}(w\,z_l^k)
    &= \prod_{T'\in{\cal T}^{\eta(k,m)}[l-1]}\ (1-w\,z_l^k\,Z'^{\,\wgt(T')})
       \ M(Z')\,.
\end{align}
It remains to take the product over all $k=1,2,\ldots,p-1$ and
$m=1,2,\ldots,m(k)$, but this is a product over all shapes $\eta(k,m)$ that
are obtained by the removal of a horizontal strip~\cite[p.5]{macdonald:1995a}
of $k$-boxes from the shape of $\pi$. For each semistandard tableau $T'$ of
shape $\eta(k,m)$ with entries from $\{1,2,\ldots,l-1\}$, if we then fill the
$k$ boxes of the horizontal strip with $k$ entries $l$ we obtain a semistandard
tableau $T$ of shape $\pi$ with entries from $\{1,2,\ldots,l\}$. All 
semistandard tableaux $T$ of shape $\pi$ containing at least one $l$ and no
more than $p-1$ entries $l$ can be obtained in this way. 

Combining this with our earlier result on the first factor, implies that
\begin{align}
 M(Z)/L_\pi(w)
   &= (\,M(z_l)/L_{(p)}(w)\,)\  (\, \prod_{k=1}^{p-1}\, 
       \prod_{m=1}^{m(k)}\ L_{\eta(k,m)}(w\,z_l^k) \,) \,)
   \nonumber \\
   &= M(z_l)\ \prod_{T\in{\cal T}^{\pi}[l]_l}\ (1-w\,Z^{\,\wgt(T)})\ 
        (\,M(Z')/L_\pi(w)\,)\,,
\end{align}
where the subscript $l$ on $[l]$ is intended to indicate that the product is
taken over all those semistandard tableaux $T$ of shape $\pi$ containing at
least one entry $l$.

By applying the same process to $M(Z')/L_\pi(w)$, one obtains factors
corresponding to all semistandard $T$ of shape $\pi$ containing no entry $l$
but at least one entry $l-1$. Continuing with this iteration procedure one
obtains the result
\begin{align}
 M(Z)/L_\pi(w) 
   &= \displaystyle  \prod_{T\in{\cal T}^{\pi}[l]}\ (1-w\,Z^{\,\wgt(T)})\ 
         M(z_l)\,M(z_{l-1})\,\cdots\,M(z_1)\,,
\end{align}
thereby completing the proof of (\ref{Eq-Lpi-perpw-M}).
\qed

It should be noted that, as one possible definition of Schur functions, 
\begin{align}
    s_\pi(Z) &= \sum_{T\in{\cal T}^{\pi}[l]}\ Z^{\,\wgt(T)}\,,
\end{align}
since the monomials in the expansion of $s_\pi(Z)$ are precisely the various
$Z^{\,\wgt(T)}$ specified by all the semistandard tableaux $T$ appearing in
${\cal T}^{\pi}[l]$. Then, thanks to the plethystic definition of 
$L_\pi(w;Z)$, (\ref{Eq-Lpi-perpw-M}) immediately implies the validity of:

\mybenv{Corollary}
\begin{align}
     L_\pi^\perp(w)\,(M(Z)) &= L_\pi(w;Z) M(Z)\,.
\end{align}
\myeenv

Once it is recalled that the dependence on $X$ has been omitted, this can be
seen to be nothing other than an exemplification of the more general
result~(\ref{Eq-FM-FperpM}) associated with the Cauchy kernel. 

%
%

\section{Proof of adjoint action identity\label{appendix:bchd}}
\setcounter{equation}{0}

\mybenv{Theorem}
\label{theorem:bchd}
Let $x$ and $y$ be arbitrary elements of a ring $R$ with identity $1$ but which is in general
non-commutative. Then
\begin{align}
\label{eq:bchd}
  \exp(x) \exp(y) \exp(-x) = \exp \left( \sum_{n=0}^\infty \frac{1}{n!} [x,\cdots,[x,[x,y]]\cdots]\right)\,,      
\end{align}
where the displayed commutator $[x,\cdots,[x,[x,y]]\cdots]$ is of degree $n$ in $x$.
\myeenv
\medskip

\noindent{\bf Proof}:\ \
For all $x\in R$ the mutully inverse functions $\exp$ and $\ln$ are defined by 
\begin{align}
    \exp(x) = \sum_{k=0}^\infty  \frac{1}{k!}\, x^k    
    &\qquad\quad\hbox{and}&  \ln(x)= \sum_{m=1}^\infty \frac{(-1)^{m-1}}{m}\, (x-1)^m\,.
\end{align}
Now let $\exp(x) \exp(y) \exp(-x) = \exp(z)$ so that
$z=\ln(\exp(x) \exp(y) \exp(-x))$. It follows that
\begin{align}
\label{eq:z}
  z &= \sum_{m=1}^\infty \frac{(-1)^{m-1}}{m}\, \left( 
                \sum_{p,q,r\geq0} (-1)^r\frac{x^py^qx^r}{p!q!r!}   
                \ -\ 1\right)^m
    \nonumber\\
   \displaybreak
    &= \sum_{m=1}^\infty \frac{(-1)^{m-1}}{m}\,
            \sum_{\begin{array}{c}p_i,q_i,r_i\geq0\cr p_i+q_i+r_i>0\cr\end{array}}
             (-1)^{r_1}\frac{x^{p_1}y^{q_1}x^{r_1}}{p_1!q_1!r_1!} 
              (-1)^{r_2}\frac{x^{p_2}y^{q_2}x^{r_2}}{p_2!q_2!r_2!} 
              \cdots
               (-1)^{r_m}\frac{x^{p_m}y^{q_m}x^{r_m}}{p_m!q_m!r_m!}\,.              
\end{align}

In this expansion as a signed sum of products of triples consider those
contributions for which a triple contains no $y$ that is to say
a triple of the form $(x^p y^0 x^r)$ with $p+r=n>0$. For each such $n$, if we 
collect together all those terms that differ only in the values of $p$ 
and $r$ their sum contains the factor
\begin{align}
    \sum_{r=0}^n (-1)^r\frac{x^{n-r}x^r}{(n-r)!r!} = \frac{1}{n!}(x-x)^n =0.
\end{align}
It follows that in (\ref{eq:z}) we need retain only those terms for which
$q_k>0$ for all $k=1,2,\ldots,m$.

Now consider those terms for which there are two neighbouring triples 
$(\cdots y x^r)(x^p y \cdots)$ with $r+p=n>0$. Then as before, for each 
such $n$, if we collect together all those terms that differ only in the 
values of $p$ and $r$ their sum contains once again the factor
\begin{align}
    \sum_{r=0}^n (-1)^r\frac{x^{n-r}x^r}{(n-r)!r!} = \frac{1}{n!}(x-x)^n =0.
\end{align}
It follows that in (\ref{eq:z}) we need retain only those terms for which
no two $y$'s are separated by any $x$'s.

This leaves only terms of the form $x^py^qx^r$, with $q>0$ and $p,r\geq0$.
As far as the constituent triples
are concerned the $x^p$ and $x^r$ must be attached to at least one $y$
on their right and left, respectively, since all triples consisting of
just $x$'s have been eliminated. Thus the contribution of the $x$'s 
is a fixed common factor, namely $(-1)^r x^{p+r}/(p!r!)$. Apart from this 
common factor the contribution of all terms $x^py^qx^r$ to $z$ in 
(\ref{eq:z}) is given by
\begin{align}
\label{eq:qn}
   &\sum_{m=1}^\infty \frac{(-1)^{m-1}}{m}\,
            \sum_{q_i>0;\ q_1+q_2+\cdots+q_m=q}\
             \frac{y^{q_1+q_2+\cdots+q_m}} 
               {q_1!q_2!\cdots q_m!} \nonumber\\
   &= \sum_{m=1}^\infty \frac{(-1)^{m-1}}{m}\,  \frac{y^q}{q!} \,
      \sum_{q_i>0;\ q_1+q_2+\cdots+q_m=q}\
       \left( \begin{array}{c} q\cr q_1\, q_2\, \ldots\, q_m\cr \end{array} \right) \nonumber\\          
   &= \sum_{m=1}^\infty \frac{(-1)^{m-1}}{m}\,  \frac{y^q}{q!}\, m!\, S(q,m)
    = \frac{y^q}{q!} \, \sum_{m=1}^\infty s(m,1)\, S(q,m) = \frac{y^q}{q!}\, \delta_{q,1}\,.              
\end{align}  
where $s(m,1)$ and $S(q,m)$ are Stirling numbers of the first and second kind, respectively. 

It follows that the only surviving terms in (\ref{eq:z}) are those of the
form $x^pyx^r$ with $p,r\geq0$, and each of these terms must constitute a single triple,
with $m=1$. Thus
\begin{align}
  z &= \sum_{p,r\geq0}\ (-1)^r \frac{x^p y x^r}{p! r!}
     = \sum_{n=0}^\infty\ \frac{1}{n!}\ \sum_{r=0}^n (-1)^r \left(\begin{array}{c} n\cr r\cr\end{array}\right) x^{n-r} y\, x^r\,.
\end{align} 

To complete the proof of Theorem~\ref{theorem:bchd} it only remain to prove the following:

\mybenv{Lemma}
\label{lemma:ncommutator}
For all $x$ and $y$ and all non-negative integers $n$
\begin{align}
\label{eq:ncommutator}
    [x,\cdots,[x,[x,y]]\cdots] =  \sum_{r=0}^n (-1)^r 
    \left(\begin{array}{c} n\cr r\cr\end{array}\right) x^{n-r}\, y\, x^r\,,
\end{align}
where the commutator on the left is of degree $n$ in $x$.
\myeenv
\medskip

\noindent{\bf Proof}:\ \ 
We offer a proof by induction with respect to $n$. For $n=0$ the right hand side
of (\ref{eq:ncommutator}) is just $y$, and this is how the left hand side
must be interpreted in this $n=0$ case. Perhaps more significantly, for $n=1$
the right hand side of (\ref{eq:ncommutator}) reduces to $xy-yx=[x,y]$, as required.

Now, for convenience, let $[x^{(k)},y]$ denote the commutator $[x,\cdots,[x,[x,y]]\cdots]$
of degree $k$ for any positive integer $k$.
Then assuming the validity of (\ref{eq:ncommutator}) in the case $n=k$ we have
\begin{align}
    [x^{(k+1)},y]&= [x,[x^{(k)},y]]\nonumber\\
          &= \sum_{r=0}^k (-1)^r \left(\begin{array}{c} k\cr r\cr\end{array}\right)  x^{k-r+1}\,y\,x^r
               -\sum_{r=0}^k (-1)^r \left(\begin{array}{c} k\cr r\cr\end{array}\right) x^{k-r}\,y\,x^{r+1}\nonumber \\
          &= x^{k+1}\,y + \sum_{r=1}^{k} \left(
           \left(\begin{array}{c} k\cr r\cr\end{array}\right) +  \left(\begin{array}{c} k\cr r-1\cr\end{array}\right) 
           \right) x^{k+1-r}\,y\,x^{r} + (-1)^{k+1}y\,x^{k+1}\nonumber \\
          &= \sum_{r=0}^{k+1} (-1)^r \left(\begin{array}{c} k+1\cr r\cr\end{array}\right)  x^{k+1-r}\,y\,x^r\,.
\end{align}
This proves the required result for $n=k+1$ and completes the induction argument, thereby proving Lemma~\ref{lemma:ncommutator} and hence also
Theorem~\ref{theorem:bchd}.
\qed

%
%
\medskip
\section{A general routine to compute scaled plethysms\label{appendix:code}}
\setcounter{equation}{0} 

In this appendix we want to give pseudo code for an algorithm to compute
plethysms with scaled arguments. Such an algorithm was implemented in the
Maple package \texttt{SchurFkt}~\cite{schurfkt:2003a}. To the best knowledge
of the authors no other computer algebra system uses this fast algorithm, so
it seems appropriate to present this method here.

We assume that we have a basis $\{u_\lambda\}$ of the ring of symmetric
functions $\Lambda(X)$ in countably many variables. We distinguish basis
monomials \verb+SymB+, terms \verb+SymT+ and polynomials \verb+SymFkt+. We
need also types for the tensor product and call this \verb+SymBxB+ for tensor
basis monomials and \verb+SymFktBxB+ for general tensor polynomials.
We also assume that we can compute the following functions for this basis:

\begin{itemize}
\item $\dim : \Lambda(X) \times R \longrightarrow R$ the dimension function
for vector spaces $V^\lambda$ having an $GL(\alpha)$ action for $\alpha\in R$.
Such vector spaces need not be irreducible. We call this map\\ 
\verb+ dim :: SymB, Ring -> Ring+
\item $\Delta : \Lambda(X) \longrightarrow \Lambda(X) \otimes \Lambda(X)$
the outer coproduct. Due to self duality this is equivalent to computing
skew products. This function is called\\
\verb+Delta :: SymB -> SymFktBxB+.\\
The fast evaluation of outer coproducts
is done using, for example, the Lascoux-Sch\"ut\-zen\-ber\-ger algorithm for skew
Schur functions, see~\cite{kohnert:2010a}.
\vfill
\newpage
\item $\delta : \Lambda(X) \longrightarrow \Lambda(X) \otimes \Lambda(X)$
the inner coproduct. 
Due to self duality this is equivalent to computing an inner product. 
This function is called\\
\verb+delta :: SymB -> SymFktBxB+.\\
The inner coproduct is computed from the Kronecker coefficients 
of inner products evaluated, for example in the Schur basis, by
the method of Robinson~\cite{robinson:1961a}.
\item We also assume that we can compute plethysms for basis monomials
$\{u_\lambda\}$, choosing our favorite method. This map is called\\
\verb+plethB :: SymB, SymB -> SymFkt+. \\
Good algorithms for plethysms in
standard bases are available~\cite{chen:garsia:remmel:1984a,kohnert:2010a}.
\end{itemize}
Let us further assume, that a symmetric function (tensor) polynomial
is stored so that we can access terms by a function \verb+listOfTerms+
and that a term is a pair (triple) consisting of a coefficient in $R$ and a basis
monomial in $\{u_\lambda\}$ (a pair of basis monomials) which we can
access by functions \verb+first+ for the coefficient and \verb+second+
(and \verb+third+) for the basis monomial(s).

We know from the properties of plethysms displayed 
in~(\ref{littlewood:plethysm}), that the plethysm is linear in the first argument but not
linear in the second argument. Our task is hence to provide a procedure
for expanding with respect to a general symmetric function in the second argument. This
reads as follows:

\lstset{numbers=left, firstnumber=auto, language=C}
\begin{lstlisting}[caption={General routine for plethysms}, label=lstPlethysm]
// Declarations of predefined functions
SymFkt    	plethysmB(SymB, SymB);
SymFktBxB  	delta(SymB), Delta(SymB);
Ring		dim(SymB,Ring);

// Declarations
SymFkt		plethysmRight(SymB,SymFkt);
SymFkt		plethysm(SymFkt, SymFkt);

// Procedures
// right non-linear expansion
plethysmRight(sMon,sPoly){
	SymB 	     	ty1; // local variables
        SymT	     	term,head;
        SymFktBxB    	coProd;
        SymFkt	     	res;
        List[SymFkt] 	tail;
        List[SymFktBxB] lstTerms;
        if zero=second(sMon) { return(sMon); };
	lstTerms := listOfTerms(sPoly);
	if #lstTerms=1 {
	  coProd := delta(sMon); //inner coproduct
          lstTerms := listOfTerms(coProd);
          res := 0;
          for term in lstTerms do{
             res := res +
	     dim(second(term),first(sPoly))
                *plethysmB(third(term),sPoly);
	  }
	  return(res);
	} else {
          head := first(lstTerms);
          tail := rest(lstTerms);
          coProd := Delta(sMon); //outer coproduct
          lstTerms := listOfTerms(coProd);
          res := 0;
          for term in lstTerms do{
	    res := res + 
                   first(term)*plethysmRight(second(term),head)
                              *plethysmRight(third(term),tail);  
	  }
          return(res);
        }
	
}

// left linearity
plethysm(sPloly1, sPoly2){
	SymT	term;
	SymFkt	res :=0;
        if sPoly1=0 or sPoly2=0 then { return 0 };
	for term in listOfTerms(sPoly1) do {
	  res := res + 
	         first(term)*plethysmRight(second(term),sPoly2);
	}
	return(res);
}
\end{lstlisting}

We end this Appendix by noting that many standard maps have
a plethystic interpretation and hence are available via the above 
algorithm. Among them are the identity map $\Id$ seen as plethysm with
$s_{(1)}$ and the antipode map $\antip$ seen as plethysms with 
$((-1)s_{(1)})$.
\end{appendix}
\medskip
\vfill
\newpage



\end{document}